\documentclass[
showpacs, twocolumn]{revtex4}
\usepackage{epsfig}
\usepackage{graphicx}

\begin{document}
\title{Nonlinear theory of solitary waves \\
associated with longitudinal particle motion in lattices:
\\ Application to longitudinal grain oscillations \\ in a dust crystal
\footnote{Preprint; to appear in \textit{European Physics Journal B}.}}
\author{I. Kourakis\footnote{
On leave from: U.L.B. - Universit\'e Libre de Bruxelles,
Physique Statistique et Plasmas C. P. 231, Boulevard du Triomphe,
B-1050 Brussels,
Belgium;
also:
Facult\'e des Sciences Apliqu\'ees - C.P. 165/81 Physique
G\'en\'erale, Avenue F. D. Roosevelt 49, B-1050 Brussels, Belgium;
\\Electronic address: \texttt{ioannis@tp4.rub.de}}
and
P. K. Shukla\footnote{Electronic address: \texttt{ps@tp4.rub.de}}}
\affiliation{Institut f\"ur Theoretische Physik IV, Fakult\"at
f\"ur Physik und Astronomie, Ruhr--Universit\"at Bochum, D-44780
Bochum, Germany}
\date{submitted: 17 dec. 2003, accepted: 6 jan. 2003}

\begin{abstract}
The nonlinear aspects of longitudinal motion of interacting point
masses in a lattice are revisited, with emphasis on the paradigm
of charged dust grains in a dusty plasma (DP) crystal. Different
types of localized excitations, predicted by nonlinear wave theories,
are reviewed and conditions for
their occurrence (and characteristics) in DP crystals are
discussed. Making use of a general formulation, allowing for an arbitrary
(e.g. the Debye electrostatic or else) analytic potential form $\phi(r)$ and
arbitrarily long site-to-site range of interactions, it is shown
that dust-crystals support nonlinear kink-shaped localized excitations
propagating at velocities above the characteristic DP lattice
sound speed $v_0$. Both compressive and rarefactive kink-type
excitations are predicted, depending on the physical parameter
values, which represent pulse- (shock-)like coherent structures for
the dust grain relative displacement.
Furthermore, the existence of breather-type localized oscillations,
envelope-modulated wavepackets and shocks is established.
The relation to previous results on atomic chains as well as to
experimental results on strongly-coupled dust layers in
gas discharge plasmas is discussed.
\end{abstract}

\maketitle

\section{Introduction}

A wide variety of linear electrostatic waves are known to
propagate in plasmas \cite{Stix, Krall}.
It is now established that the inherent
nonlinearity of electrostatic dispersive media gives birth to
remarkable new phenomena, in particular related to the formation and
stable propagation of
long-lived nonlinear structures,
when a balance between nonlinearity and dispersion is possible
\cite{Karpman, Debnath}. Since about a decade ago, plasma wave
theories have received a new boost after the prediction (and
subsequent experimental confirmation) of the existence of new
oscillatory modes, associated with charged dust-grain motion in
dust-contaminated plasmas, as well as the possibility for an
important modification of existing modes due to the presence of
charged dust grains \cite{PSbook, Verheest}. A unique new feature
associated to these dusty (or complex) plasmas (DP) is the existence of
new strongly-coupled charged matter configurations, held
responsible for a plethora of new phenomena e.g. phase
transitions, crystallization, melting etc., and possibly even
leading to the formation of dust-layers (DP crystals) when the
inter-grain potential energy far exceeds the average dust kinetic
energy; a link has thus been established between plasma physics and
solid state physics \cite{Kittel}. These dust Bravais-type
quasi-lattices, which are typically formed in the sheath region in
low--temperature dusty plasma discharges,
and remain suspended above the negative electrode due
to a balance between the electric and gravity forces \cite{Chu,
Thomas, Melzer, Hayashi}, are known to support harmonic
excitations (acoustic modes) in both longitudinal and
transverse-shear (horizontal-plane) directions, as well as
optical-mode-like oscillations in the vertical (off-plane)
direction \cite{Melandso, PKSPLA2002, farokhi, vladimirov1, tskhakaya, Wang,
Morfill, Nunomura2}.

The longitudinal dust-lattice waves (LDLW) are reminiscent of
waves (`phonons') propagating in atomic chains, which
are long known to be dominated by nonlinear
phenomena, due to the intrinsic nonlinearities of inter-atomic
interaction mechanisms and/or on-site substrate potentials \cite{Tsurui,
Wadati, Pnevmatikos1986, Flytzanis1985, Wattis}. These phenomena have been
associated with a wealth of phenomena, e.g. dislocations in
crystals, energy localization, charge and information transport in
bio-molecules and DNA strands, coherent signal transmission in
electric lines, optical pulse propagation and many more
\cite{Davydov, Remo, Newell2, Scott, Peyrard}. Even though certain
well-known nonlinear mechanisms, e.g. shock formation,
electrostatic pulse propagation and instabilities,
have been thoroughly investigated in
weakly-coupled (gas-like) dusty plasmas \cite{PSbook,
PKS2003, IKPKSPoP}, the theoretical
investigation of the relevance of such phenomena with waves in
DP crystals is
still in a pre-mature stage; apart from the pioneering works of
Melands\o \cite{Melandso}, who first derived a Korteweg-DeVries (KdV)
equation \cite{Stenflo} associated with longitudinal dust-lattice oscillations, Shukla
\cite{PKSPLA2002},
who predicted the formation of dust cavitons due to lattice dynamical
coupling to surrounding ions,
and the investigation of related nonlinear amplitude
modulation effects by Amin \textit{et al.} \cite{AMS2} a little later,
not much has been done in the direction of
a systematic elucidation
of the relevance of
dust-lattice waves being described by the
known model nonlinear wave equations.
It should, however, be stressed that some recent
attempts to trace the signature of nonlinearity in experiments
\cite{Nunomura, Nosenko, Samsonov} have triggered an effort to
interpret these results in terms of coherent structure propagation
\cite{Samsonov, Nunomura, Avinash, Zhdanov}, essentially along the
physical ideas suggested in Ref. \cite{Melandso}.

In this paper, we aim at reviewing the procedure employed in the
derivation of a nonlinear evolution equation for longitudinal dust grain
motion in DP lattices, and discussing the characteristics of the
solutions. Emphasis is made on the methodology, in a quite exhaustive
manner, in close relation with previous results on atomic chains,
yet always focusing on the particular features of DP crystals; we
will discuss, in particular:

- the physical assumptions underlying the continuum approximation;

- the choise of truncation scheme, when departing from the
discrete lattice picture;

- the long-range electrostatic interactions, differentiating DP crystals
from ordinary classical atomic chains (spring models);

- the physical relation between different solutions obtained.
\newline
Some of the results presented here are closely related to
well-known previous results, yet enriched with a new analytical
set of coefficients allowing for any assumed range of site-to-site
interactions and any analytical form of the interaction potential.
The present
study is, therefore, valid in both short and long- Debye length DP
cases, and also aims at providing a general `recipe' which allows
one, for instance, to assume a modified (possibly non-Debye-type)
potential form and obtain the corresponding set of formulae in a
straightforward manner. In specific, we have in mind the
modification of the inter-grain interactions due to ion flow in
the sheath region surrounding the dust layer, which may even lead
to the crystal being destabilized, according to recent studies from first
principles \cite{IKPKSPLA, Ignatov}.

Most of the results presented here are general and apply, in principle,
to a sufficiently general class of chains of classical agents
(point masses) coupled via arbitrary (and possibly long-range) interaction
laws.
Nevertheless, our specific aim is to establish a first link between
existing nonlinear theories and the description of
longitudinal dust-lattice oscillatory grain motion in a DP crystal.
At a first step, our description cannot help being
`academic', and somewhat abstract:
an
ideal one--dimensional DP crystal is considered,
i.e. a single, unidimensional, infinite-sized,
dust-layer of identical (in size, charge and mass) dust grains
situated at spatially periodic sites (at equilibrium).
Effects associated with crystal asymmetries, defects, dust charging,
ion-drag,
dust mass variation and multiple dust-layer coupling, are left for
further consideration \cite{Ivlev}.
Transverse (off-plane) motion, in particular, will be addressed in
a future work.

\section{The model}

\subsection{Equation of motion}

Let us consider a layer of charged dust grains
(mass $M$, charge $Q$, both assumed constant
for simplicity) forming a Bravais lattice, of lattice constant $r_0$.
The Hamiltonian of such a chain reads
\[H = \sum_n \frac{1}{2} \, M \, \biggl( \frac{d \mathbf{r}_n}{dt} \biggr)^2 \, + \,
\sum_{m \ne n} U(r_{nm}) \, ,
\]
where $\mathbf{r}_n$ is the position vector of the $n-$th grain;
$U_{nm}(r_{nm}) \equiv Q \, \phi(x)$ is a binary interaction
potential function related to the electrostatic potential
$\phi(x)$ around the $m-$th grain, and $r_{nm} =
|\mathbf{r}_{n}-\mathbf{r}_{m}|$ is the distance between the
$n-$th and $m-$th grains. We shall limit ourselves to considering
the {\em longitudinal} ($\sim \hat x $) motion of the $n-$th dust
grain, which obeys
\begin{equation}
M\, \biggl( \frac{d^2 x_n}{dt^2} \, + \nu \,
\frac{d x_n}{dt}
\biggr)
= \,- \sum_n \,\frac{\partial U_{nm}(r_{nm})}{\partial x_n}
\, \equiv \, Q
\, E(x_n),
\label{eqmotion0}
\end{equation}
where $E(x) = - \partial \phi(x)/\partial x$ is the electric
field; the usual \textit{ad hoc} damping term is introduced in the
left-hand-side (lhs), involving the damping rate $\nu$, to account
for the dust grain collisions with neutrals. Note that a
one-dimensional (1D) DP layer is considered here, but the
generalization to a two-dimensional (2D) grid is straightforward.
At a first step, we have omitted the external force term
$F_{ext}$, often introduced to account for the initial laser
excitation and/or the parabolic confinement which ensures
horizontal lattice equilibrium in experiments \cite{Samsonov}.
The analogous formulae for non-electrostatic, e.g. spring--like coupling
interactions are readily obtained upon some trivial modifications
in the notation.

The additive structure of the contribution of each site to the
potential interaction force in the
right-hand-side (rhs) of Eq. (\ref{eqmotion0}) allows us to express
the electric field in (\ref{eqmotion0}) as:
\begin{eqnarray}
E(x) & = & \,- \frac{\partial }{\partial x_n} \sum_{m} \,
\phi(x_n - x_{m})
\nonumber \\
& = & \, + \sum_{l} \,
\bigl[\phi'(x_{n+l} - x_{n}) - \phi'(x_n - x_{n-l}) \bigr]
\nonumber \\
& = &
\sum_{l = 1}^N \,\sum_{l'=1}^\infty \,\frac{1}{l'!}
\biggl. \frac{d^{l'+1} \phi(r)}{d r^{l'+1}} \biggr|_{r = l r_0} \,\times
\nonumber \\ & &
\bigl[(\delta x_{n+l} - \delta x_{n})^{l'} -
(\delta x_n - \delta x_{n-l})^{l'} \bigr]
\end{eqnarray}
where $l$ denotes the degree of vicinity, i.e.
$l = 1$ accounts for the nearest-neighbour interactions (NNI) and
$l \ge 2$ accounts for distant- (second or farther) neighbour
interactions (DNI). The summation upper limit $N$
naturally depends on the model
and the interaction mechanism; even though $N$ `traditionally'
equals either 1 or 2 in most studies of atomic chains, one should
consider higher values for long-range-interactions e.g. Coulomb or Debye
(screened) electrostatic interactions (the latter case is addressed below,
in detail).
In the last step, we have Taylor-developed
the interaction potential $\phi(r)$ around the equilibrium inter-grain
distance $l r_0 = |n-m| r_0$ (between $l-$th order neighbours), viz.
\[\phi(r_{nm}) = \sum_{l'=0}^\infty \,\frac{1}{l'!}
\biggl. \frac{d^{l'} \phi(r)}{d r^{l'}} \biggr|_{r = |n-m| r_0} \,
(x_{n} - x_{m})^{l'} \, ,\]
where $l'$ denotes
the degree (power) of nonlinearity involved in each contribution:
$l' = 1$ is the
linear interaction term, $l' = 2$ stands for the
quadratic nonlinearity, and so forth.
Obviously, $\delta x_n =  x_n -  x_n^{(0)}$
denotes the displacement of the
$n-$th grain from equilibrium, which now obeys
\begin{eqnarray}
M\, \biggl( \frac{d^2 (\delta x_n)}{dt^2} \, + \nu \, \frac{d
(\delta x_n)}{dt} \biggr) = \qquad \qquad \qquad \qquad  \nonumber \\
\, Q \biggl\{ \phi''(r_0) \, (\delta
x_{n+1} + \delta x_{n-1} - 2 \delta x_{n}) \qquad \nonumber \\
\, + \sum_{l=2}^N \,\phi''(l r_0) \, (\delta x_{n+l} +
\delta x_{n-l} - 2 \delta x_{n})
\nonumber \\
\, + \sum_{l'=2}^\infty \frac{1}{l'!}\,\biggl. \frac{d
\phi^{l'+1}(r)}{d r^{l'+1}} \biggr|_{r = r_0}  \, \times  \qquad \qquad
\qquad
 \nonumber \\
 \bigl[ (\delta
x_{n+1} - \delta x_{n})^{l'} -  (\delta x_{n} - \delta
x_{n-1})^{l'} \bigr]
 \nonumber \\
\, + \sum_{l = 2}^N \sum_{l' = 2}^\infty
\frac{1}{l'!}\,\biggl. \frac{d \phi^{l'+1}(r)}{d r^{l'+1}}
\biggr|_{r = l r_0}  \, \times \qquad  \qquad  \nonumber \\
\biggl[ (\delta x_{n+l} - \delta
x_{n})^{l'} -  (\delta x_{n} -
\delta x_{n-l})^{l'} \biggr] \biggr\}  \, .\nonumber \\
 \label{eqmotion1}
\end{eqnarray}
We have distinguished the linear/nonlinear contributions of the
first neighbors (1st/3rd lines) from the corresponding longer
neighbor terms (2nd/4th lines, respectively).

Keeping all upper summation limits at infinity, the last discrete
difference equation (\ref{eqmotion1}) is exactly equivalent to
the complete equation (\ref{eqmotion0}).
However, the former needs to be truncated to a specific order in $l, l'$,
depending on the desired level of sophistication, for reasons of
tractability.

\subsection{Continuum approximation.}

We shall now adopt the standard continuum approximation often
employed in solid state physics \cite{Kittel}, trying to be
very systematic and keeping track of any inevitable term truncation.
We will
assume that only small displacement variations occur between
neighboring sites, i.e.
\[
\delta x_{n \pm l} \, = \, \delta x_{n} \, \pm \, l r_0 \frac{\partial
u}{\partial x} \, + \, \frac{1}{2} (l r_0)^2 \frac{\partial^2
u}{\partial x^2} \, \qquad \qquad \qquad \]
\[ \qquad \qquad\qquad\qquad
\pm \, \frac{1}{3!} (l r_0)^3 \frac{\partial^3
u}{\partial x^3} \,
+ \, \frac{1}{4!} (l r_0)^4 \frac{\partial^4
u}{\partial x^4} + ... \ ,
\]
i.e.
\begin{eqnarray}
\delta x_{n + l} - \delta x_{n} = l r_0 \frac{\partial
u}{\partial x} + \frac{1}{2} (l r_0)^2 \frac{\partial^2
u}{\partial x^2} \qquad\qquad\qquad\nonumber \\
\qquad + \frac{1}{3!} (l r_0)^3 \frac{\partial^3
u}{\partial x^3} + \frac{1}{4!} (l r_0)^4
\frac{\partial^4 u}{\partial x^4} + ... \nonumber \\
= \sum_{m=1}^\infty \frac{(l r_0)^m}{m!} \, \frac{\partial ^m
u}{\partial x^m} \ , \qquad\qquad\qquad\label{diff1}
\end{eqnarray}
and \begin{eqnarray} \delta x_{n} - \delta x_{n-1} = l r_0
\frac{\partial u}{\partial x} - \frac{1}{2} (l r_0)^2
\frac{\partial^2 u}{\partial x^2}  \qquad\qquad\qquad
\nonumber \\
 \qquad
\qquad+ \frac{1}{3!} (l r_0)^3
\frac{\partial^3 u}{\partial x^3} - \frac{1}{4!} (l r_0)^4
\frac{\partial^4 u}{\partial x^4} + ... \nonumber \\
= - \sum_{m=1}^\infty (-1)^m \,\frac{(l r_0)^m}{m!} \,
\frac{\partial ^m u}{\partial x^m} \, ,
\qquad
 \label{diff2}
\end{eqnarray}
 where the displacement $\delta x (t)$ is now expressed as a
continuous function $u = u(x, t)$.

Accordingly, the linear contributions (i.e. the first two lines)
in (\ref{eqmotion1}) now give
\begin{eqnarray}
Q \sum_{l = 1}^N \phi''(l r_0)\,  \, (\delta x_{n+l} + \delta
x_{n-l} - 2 \delta x_{n}) \, \qquad\qquad \nonumber \\
= \, Q \, \sum_{m = 1}^\infty
\sum_{l = 1}^N
\phi''(l r_0)\, (l r_0)^{2m}\,\frac{2}{(2m)!} \,
\frac{\partial^{2m}
u}{\partial x^{2m}} \nonumber \\
 =  \, Q \, \sum_{m = 1}^\infty \frac{2}{(2m)!} \, \biggl[
\sum_{l = 1}^N \phi''(l r_0)\, (l
r_0)^{2m}\biggr]\,\frac{\partial^{2m} u}{\partial
x^{2m}} \nonumber \\
 \equiv   M\,\sum_{m = 1}^\infty \, c_{2 m}\,\frac{\partial^{2m}
u}{\partial x^{2m}} \qquad \qquad\qquad\qquad\qquad \nonumber \\
 =  M\, \biggl( c_{2}\,u_{xx} + c_{4}\,u_{xxxx} +
c_{6}\,u_{xxxxxx}+ \, ...\biggr) \,  \label{linear-detailed}
\end{eqnarray}
where the subscript in $u_x$ denotes differentiation with
respect to $x$, i.e. $u_{xx} = {\partial^{2} u}/{\partial x^{2}}$
and so on. We see that only even order derivatives contribute to
the linear part; this is rather expected, since the model (for
$\nu = 0$) is conservative, whereas odd-order derivatives might
introduce a dissipative effect, e.g. via a Burgers-like ($\sim
u_{xx}$) additional term in the KdV Eq. below \cite{Grimshaw,
PKSAM, Melandsoshocks}. The definition of the coefficients $c_{2
m}$ ($m=1, 2, ...$) is obvious; the first term reads
\begin{equation} c_2 = \frac{Q}{M} \,r_0^{2}
\sum_{l = 1}^N \phi''(l r_0)\, l^2 \, \equiv v_0^2 \, \equiv
\, \omega_{0, L}^2 \,r_0^2 \, ,
\label{defv0}
\end{equation}
which defines the characteristic second-order dispersion
(`sound') velocity $v_0$ [cf. $v_p$ in (6)
of Ref. \cite{Avinash}], related to the longitudinal oscillation
eigenfrequency $\omega_{0, L}$; also
\[c_4 = \frac{1}{12} \frac{Q}{M}
\,r_0^{4} \sum_{l = 1}^N \phi''(l r_0)\, l^4 \,\equiv \,
v_1^2 \, r_0^2 ,
\]
\begin{equation}
c_6 = \frac{2}{6!} \frac{Q}{M} \,r_0^{6}
\sum_{l = 1}^N \phi''(l r_0)\, l^6 \, ,
\label{defv1}
\end{equation}
and so on. Notice that $v_1^2 = v_0^2/12$ for
NNI, i.e. if (and only if) one stops the summation at $l_{max}= N = 1$,
like Eq. (26) in Ref. \cite{Melandso}
[and \emph{unlike} Eq. (5) in Ref. \cite{Avinash},
whose 2nd term in the rhs is rather not correct,
for $l \ne 1$ i.e. DNI].
See that the `relative
weight' of any given $2m-$th contribution as compared to the
previous one is roughly $(2m - 2)!/(2 m)!$, e.g. $4!/6! = 1/30$
for $m = 3$, which somehow justifies higher (than, say, $m = 2$)
order contributions often neglected in the past;
nevertheless, this argument should rigorously not be taken for
granted, as a given function $u(x, t)$ and/or potential $\phi(x)$,
may present higher numerical
values of higher-order derivatives, balancing this numerical
effect; clearly, any truncation in an infinite series
inevitably implies loss
of information.

We may now treat the quadratic nonlinearity contribution in
(\ref{eqmotion1}) (the last two lines for $l'=2$) in the same manner.
Making use of Eqs. (\ref{diff1}) and (\ref{diff2}),
and also of the
identity $a^2-b^2 = (a+b)(a-b)$, one obtains
\[ Q \, \frac{1}{2!}
\sum_{l = 1}^N \phi'''(l r_0)\, \biggl[ (\delta x_{n+l} -
\delta x_{n})^2 - (\delta x_{n} - \delta x_{n-1})^2 \biggr] \,
\qquad \qquad \qquad \qquad
\]
\begin{eqnarray}
\qquad &=& \, Q \, \sum_{m = 1}^\infty  \sum_{m' = 1}^\infty  \,
\frac{2}{(2m-1)! (2 m')!} \, \times \,
\nonumber \\
& &
\biggl[ \sum_{l = 1}^N \phi'''(l
r_0)\,(l r_0)^{2(m+m')-1} \biggr]\, \frac{\partial^{2m-1}
u}{\partial x^{2m-1}}\,
\frac{\partial^{2m'} u}{\partial x^{2m'}} \nonumber \\
&\equiv &   M\,\sum_{m = 1}^\infty \sum_{m' = 1}^\infty \, c_{m, m'}\,
\frac{\partial^{2m-1} u}{\partial x^{2m-1}}\,
\frac{\partial^{2m'} u}{\partial x^{2m'}} \nonumber \\
&= & M\, \biggl( c_{1, 1}\,u_{x}\,u_{xx}\, + c_{1,
2}\,u_{x}\,u_{xxxx}\,
\nonumber \\
&  & \qquad \qquad \qquad \qquad
+ c_{2, 1}\,u_{xx}\,u_{xxx}\,+ \, ...\biggr)\, .
\label{quadratic-detailed}
\end{eqnarray}
The definition of the coefficients $c_{m, m'}$ ($m, m' = 1, 2,
...$) is obvious; the first few terms read
\begin{equation}  c_{1, 1} = \frac{Q}{M} \,r_0^{3}
\sum_{l = 1}^N \phi'''(l r_0)\, l^3 \, ,
\label{defc11}
\end{equation}
which defines the first nonlinear contribution [eg. $B$ in Eqs. (5)
and (7) in Ref. \cite{Avinash};
we note that a factor $1/2$ and $1/M$ is missing
therein, respectively],
\[ c_{1, 2} = \frac{2}{1! \,4!} \frac{Q}{M} \,r_0^{5}
\, \sum_{l = 1}^N \phi'''(l r_0)\, l^5 \, , \]
\[c_{2, 1} =
\frac{2}{3! \, 2!} \frac{Q}{M} \,r_0^{5} \, \sum_{l = 1}^N
\phi'''(l r_0)\, l^5 \, ,
\]
and so on.

The cubic nonlinearities in (\ref{eqmotion1})
(the last two lines for
$l'=3$) may now be treated in the same manner. Making use of
Eqs. (\ref{diff1}) and (\ref{diff2}),
as well as of the identity: $a^3 - b^3 = (a
- b)(a^2 + a b + b^2)$, one obtains
\[ Q \, \frac{1}{3!}
\sum_{l = 1}^N \phi'''(l r_0)\, \biggl[ (\delta x_{n+l} -
\delta x_{n})^3 - (\delta x_{n} - \delta x_{n-1})^3 \biggr] \,
\qquad \qquad \qquad \qquad \qquad \qquad \qquad \qquad
\]
\begin{eqnarray}
&=& \, Q \, \frac{1}{3}\, \sum_{m = 1}^\infty  \sum_{m' =
1}^\infty \sum_{m'' = 1}^\infty \, \frac{1 - (-1)^{m'} +
(-1)^{m'+m''}}{(2m)! \, m'! \, m''!} \, \times
\nonumber \\
& &
\biggl[ \sum_{l =
1}^N \phi''''(l r_0)\,(l r_0)^{2 m + m' + m''} \biggr]\,
\frac{\partial^{2m} u}{\partial x^{2m}}\, \frac{\partial^{m'}
u}{\partial x^{m'}} \,
\frac{\partial^{m''} u}{\partial x^{m''}} \nonumber \\
&\equiv &   M\,\sum_{m = 1}^\infty \, c_{m, m', m''}\,
\frac{\partial^{2m} u}{\partial x^{2m}}\, \frac{\partial^{m'}
u}{\partial x^{m'}} \,
\frac{\partial^{m''} u}{\partial x^{m''}}\nonumber \\
&= & M\, \biggl[ c_{1, 1, 1}\,(u_{x})^2\,u_{xx}\, + ( c_{1, 1, 2}
+ c_{1, 2, 1}) \, u_{x}\,(u_{xx})^2\,
\nonumber \\
& &
 \qquad \qquad \qquad \qquad
 + \, c_{1, 2, 2
}\,(u_{xx})^3\,+ \, ...\biggr] \, .\label{cubic-detailed}
\end{eqnarray}
The definition of the coefficients $c_{m, m', m''}$ ($m, m', m'' =
1, 2, ...$) is obvious; their form is immediately deduced upon
inspection, e.g.
\[ c_{1, 1, 1} = \frac{1}{2} \frac{Q}{M} \,r_0^{4}
\, \sum_{l = 1}^N \phi''''(l r_0)\, l^4 \, ,
\]
\begin{equation} \qquad c_{1, 1,
2} = - c_{1, 2, 1} = \frac{1}{12} \frac{Q}{M} \,r_0^{5} \, \sum_{l
= 1}^N \phi''''(l r_0)\, l^5 \, ,
\label{defc111}
\end{equation}
and so forth.
We note that the second term in Eq. (\ref{cubic-detailed})
cancels.

Higher order nonlinearities in Eq. (\ref{eqmotion1})
(the last two lines therein
for $l' \ge 4$), related to fifth- (or higher-) order derivatives
of the interaction potential $\phi$, will deliberately be
neglected in the following, since they are rather not likely to
affect the dynamics of small grain displacements. It should be
pointed out that, rigorously speaking, there is no \textit{a priori} criterion of
whether some truncation of the above infinite sums is preferable
to another; some \textit{ad hoc} truncation schemes, proposed in
the past, should only be judged upon by careful numerical comparison
of the relevant contributions -- e.g. in Eqs. (\ref{linear-detailed}),
(\ref{quadratic-detailed}), (\ref{cubic-detailed}) above --
and/or, finally, a comparison of the analytical results derived to
experimental ones.

Keeping the first few contributions in the above sums, one obtains
the continuum analog of the discrete equation of motion
\begin{equation}
\ddot{u}  \,+ \, \nu \, \dot{u} - v_0^2 \, u_{xx}\,= \, v_1^2\,
r_0^2 \, u_{xxxx}\, - \, p_0 \, u_x \,u_{xx} \, + \, q_0 \, (u_x)^2 \,u_{xx}
\, ,
\label{eqmotion-gen-continuum}
\end{equation}
which is the final result of this section.
Notice that $u_x \, u_{xx} = (u_x^2)_x/2$; also,
$(u_x)^2 \, u_{xx} = (u_x^3)_x/3$.
The coefficients \[v_0^2 \equiv c_2 \, , \qquad
v_1^2 \, r_0^2 \,\equiv c_4 \, ,
\qquad p_0 \equiv - c_{1, 1} \, , \qquad q_0 \equiv c_{1, 1, 1} \, , \]
defined by Eqs. (\ref{defv0}), (\ref{defv1}), (\ref{defc11}) and
(\ref{defc111}),
respectively, should be evaluated for a given potential function $\phi$,
by truncating, if inevitable, all summations therein to a given order
$l_{max}$.
Note that, quite surprisingly,
the infinite neighbour contributions may be exactly
summed up, in the case of Debye (screened) electrostatic interactions,
as we shall show below.
Let us point out that
Eq. (\ref{eqmotion-gen-continuum}) is general;
the only assumption made is the
continuum approximation.
Also, should one prefer to improve the above
truncation scheme, e.g. by including more nonlinear terms, one may readily
go back to the above formulae and simply keep one or more extra term(s);
in any case,
one can find the exact form of all (retained and truncated)
coefficients above.
On the other hand, Eq. (\ref{eqmotion-gen-continuum}) generalizes
the previous
known results for monoatomic lattices in that it holds for an arbitrary
degree of inter-site vicinity (range of interactions).

Let us point out that the above  definitions of the
coefficients in Eq. (\ref{eqmotion-gen-continuum}) are inspired by the
Debye--H\"uckel (Yukawa) potential form (whose odd/even derivatives
are negative/positive; see below), in which case they are defined in such a way that
all of $v_0^2$, $v_1^2$, $p_0$ and $q_0$ take {\em{positive}}
values. Nevertheless, keep in mind that the sign of these
coefficients for a
different potential function $\phi$ is, in principle, not
prescribed; indeed, analytical and numerical studies of the nature
of the inter-grain interactions from first principles suggest that the
presence of ion flow, for instance, may result in a structural
change in the form of $\phi$, leading to lattice oscillation
instability and presumably crystal melting \cite{Ivlev-melting};
see e.g. Refs.
\cite{Ignatov, IKPKSPLA}. However, our physical problem loses its
meaning once this happens;
therefore, we will assume, as a working
hypothesis in the following, that $c_2$ and $c_4$ bear positive
values (so that $v_0$, $v_1$ are real) - as a requirement for
the stability of the lattice -
and that, in principle (yet not necessarily),
the same holds for $p_0$ and $q_0$.

We observe that, upon setting $\nu = 0$, $q_0 = 0$, $r_0 = a$
and $l = 1$ (NNI), which imply that $v_1^2 = v_0^2 a^2/12$ and
$p_0 = - Q a^3 \phi'''(a)/M \equiv \gamma(a) a^3/M$ in
Eq. (\ref{eqmotion-gen-continuum}), one recovers exactly
Eq. (26) in Ref. \cite{Melandso}
[also see the definition in Eq. (16) therein];
also cf. Ref. \cite{AMS2}.
Equations (5) -- (7) in Ref.
\cite{Avinash} are also recovered.

In the following, we will drop the damping term
[second term in
the right-hand-side of Eq. (\ref{eqmotion-gen-continuum})], which is purely
phenomenological; the damping effect may then be re-inserted in
the analysis at any step further, by plainly adding a similar
\textit{ad hoc} term to the equation(s) modeling the grain dynamics.
It may be noted that damping comes out to be weak, in experiments
\cite{Samsonov}, so one may in pinciple proceed by including
dissipation effects \textit{a
posteriori}, and then comparing theoretical or numerical results to
experimental ones.

\subsection{An exactly computable case - the Debye ordering
\label{DebyeDPC}}

Most interestingly, the summations (in $l$) in the above
definitions of coefficients $c_{m, m', ...}$ above, converge and
may exactly be computed in the Debye-H\"uckel (Yukawa) potential
case: $\phi_{D}(r) = Q e^{-r/\lambda_D}/r$, for any given number
$N$ of neighboring site vicinity: $N = 1$ for the nearest neighbor
interactions (NNI), $N = 2$ for the second-neighbor interactions (SNI)
and even $N$ equal to infinity, for an infinite chain. The details
of the calculation are given in the Appendix, so only the final
result will be given here, for later use in this text. Note the
definition of the {\em{lattice parameter}} $\kappa = r_0/\lambda_D$, to
be extensively used in the following; in fact, $\kappa$ is roughly of the
order of (or slightly above) unity in laboratory experiments.

Truncating the summations at $N = 1$ (NNI), relations
(\ref{defv0}), (\ref{defv1}), (\ref{defc11}) and (\ref{defc111})
give
\[
({\omega_{L, 0}^{(NNI)}})^2 = \frac{2 Q^2}{M \lambda_D^3} \,
e^{-\kappa}\, \frac{1 + \kappa + \kappa^2/2}{\kappa^3}\, \]
\begin{equation}
= \,
({v_0^{(NNI)}})^2/(\kappa^2 \lambda_D^2) = 12
({v_1^{(NNI)}})^2/(\kappa^2 \lambda_D^2) \, ,
\label{om0NNI}
\end{equation}
\begin{equation}
p_0^{(NNI)} = \frac{6 Q^2}{M \lambda_D} \, e^{-\kappa}\, \biggl(
\frac{1}{\kappa} +  1 + \frac{\kappa}{2}+ \frac{\kappa^2}{6}
\biggr)\, ,
\end{equation}
\begin{equation}
q_0^{(NNI)} = \frac{12 Q^2}{M \lambda_D} \, e^{-\kappa}\, \biggl(
\frac{1}{\kappa} +  1 + \frac{\kappa}{2}+ \frac{\kappa^2}{6} +
\frac{\kappa^3}{24} \biggr)\, .
\end{equation}
These relations coincide with the ones in previous studies for
NNI \cite{Melandso, AMS2}.

Truncating the summations at $N = 2$ (SNI), relations
(\ref{defv0}), (\ref{defv1}), (\ref{defc11}) and (\ref{defc111})
give
\[
({\omega_{L, 0}^{(SNI)}})^2 = \frac{2 Q^2}{M \lambda_D^3} \,
\biggl( e^{-\kappa}\, \frac{1 + \kappa + \kappa^2/2}{\kappa^3} +
e^{-2 \kappa}\, \frac{\frac{1}{2} + \kappa + \kappa^2}{\kappa^3}
\biggr) \, \]
\begin{equation}= \, {v_0^{(SNI)}}^2/(\kappa^2 \lambda_D^2) \, ,
\end{equation}
accompanied by an extended set of expressions for
${v_1^{(SNI)}}^2$ ($\ne ({v_0^{(SNI)}})^2/12$, now, unlike in the NNI
case above), $p_0^{(SNI)}$ and $q_0^{(SNI)}$ (see in the Appendix for
details).

\begin{figure}[htb]
 \centering
 \resizebox{3in}{!}{
 \includegraphics[]{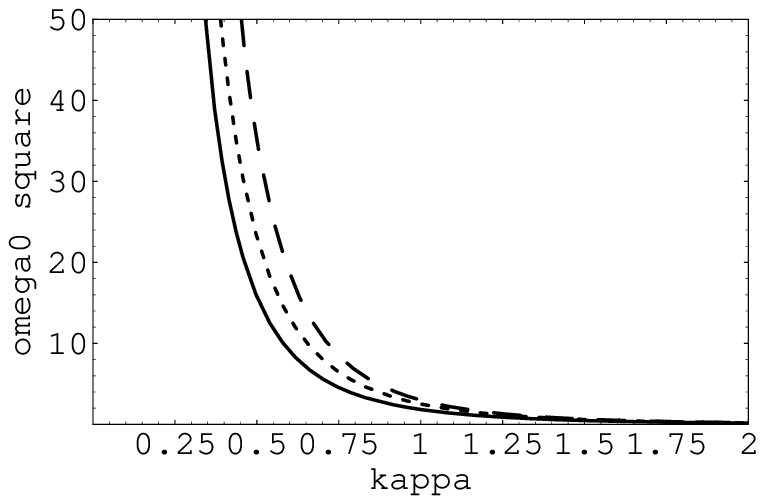}} \\
\vskip 2 cm \resizebox{3in}{!}{
 \includegraphics[]{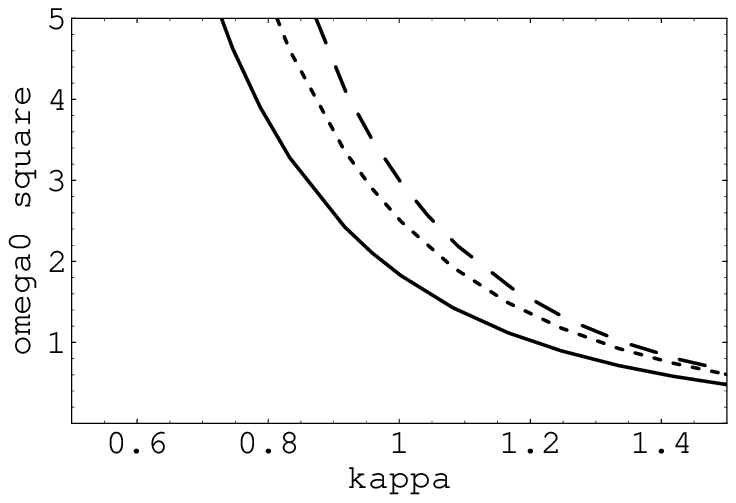}}
\caption{} \label{figure1}
\end{figure}

\begin{figure}[htb]
 \centering
 \resizebox{3in}{!}{
 \includegraphics[]{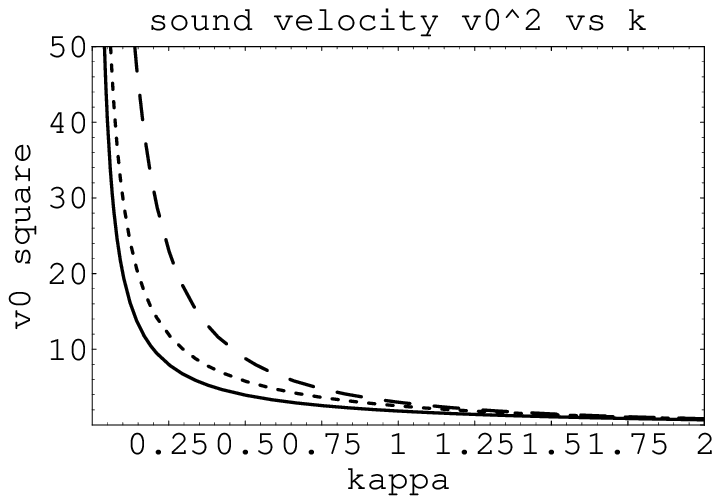}} \\
\vskip 2 cm \resizebox{3in}{!}{
 \includegraphics[]{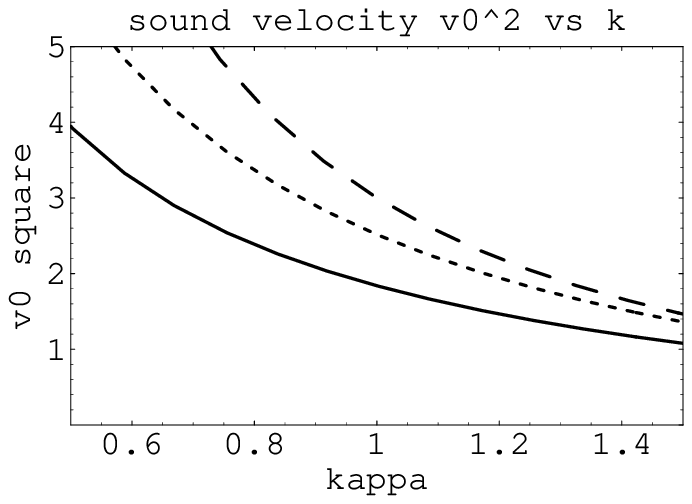}}
\caption{} \label{figure2}
\end{figure}

\begin{figure}[htb]
 \centering
 \resizebox{3in}{!}{
 \includegraphics[]{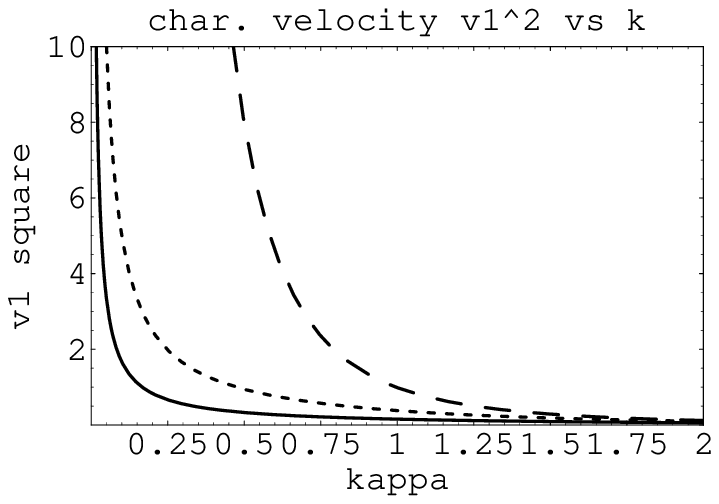}} \\
\vskip 2 cm \resizebox{3in}{!}{
 \includegraphics[]{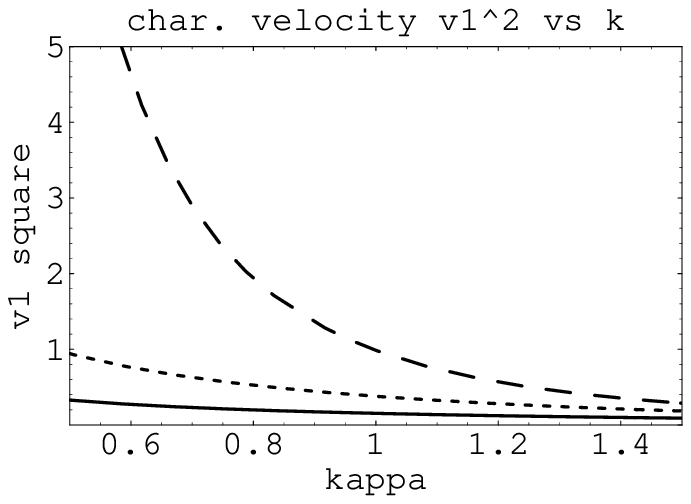}}
\caption{} \label{figure3}
\end{figure}

\begin{figure}[htb]
 \centering
 \resizebox{3in}{!}{
 \includegraphics[]{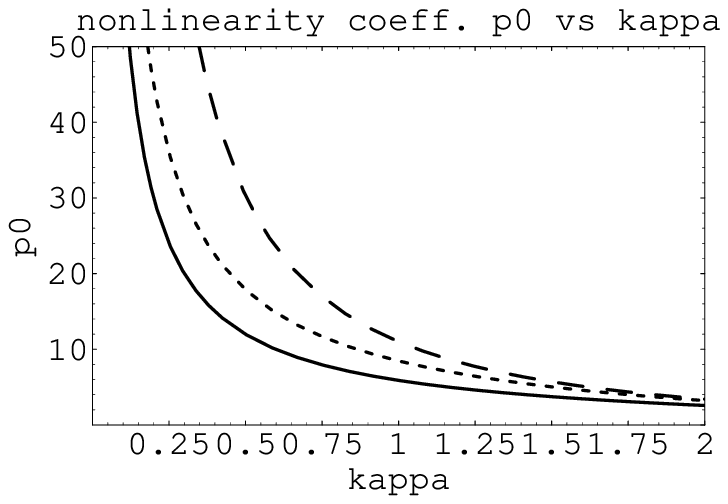}} \\
\vskip 2 cm \resizebox{3in}{!}{
 \includegraphics[]{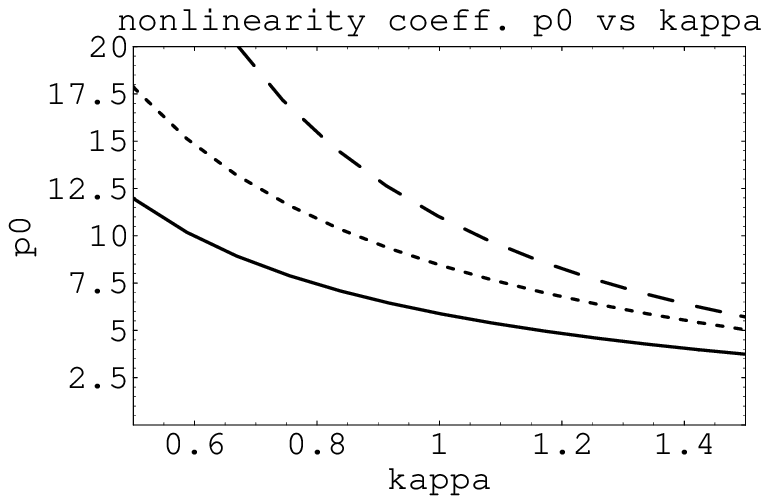}}
\caption{} \label{figure4}
\end{figure}

\begin{figure}[htb]
 \centering
 \resizebox{3in}{!}{
 \includegraphics[]{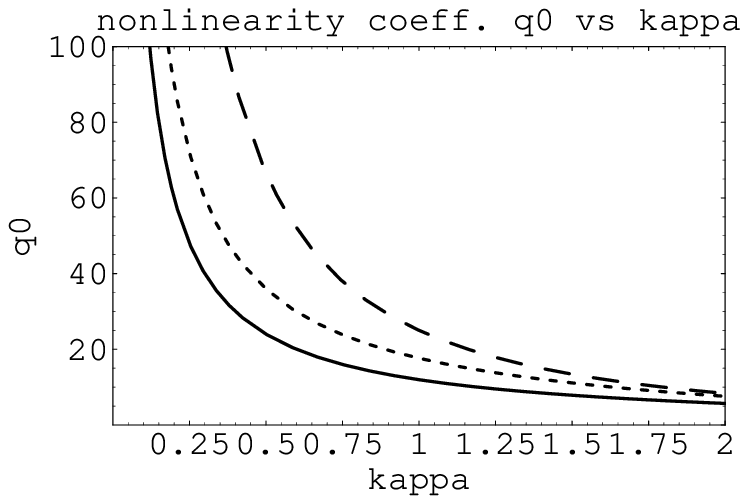}} \\
\vskip 2 cm \resizebox{3in}{!}{
 \includegraphics[]{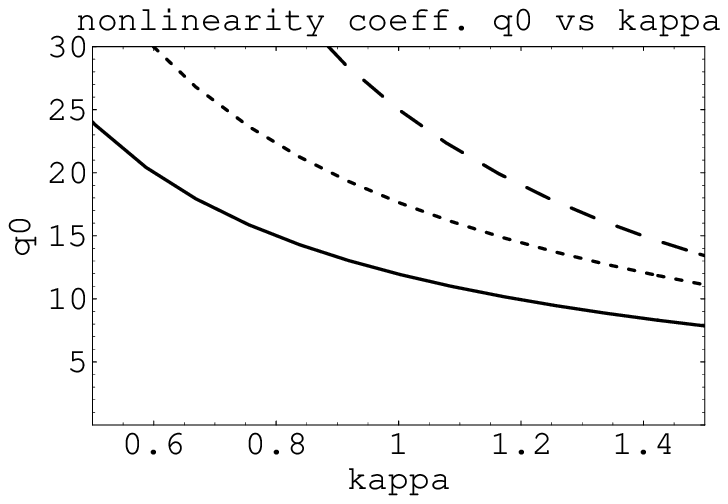}}
\caption{} \label{figure5}
\end{figure}

For higher $l_{max} = N$, even though the effect of adding more
neighbors is cumulative, since all extra contributions are
positive, these diminish fast and converge, for infinite $N$, to a
finite set of expressions, which can be calculated via the
identities: $\sum_{l = 1}^\infty a^l\, = {a}/{(1-a)}$ and $\sum_{l
= 1}^\infty {a^l}/{l} \, =  - ln(1-a)$ (for $0 < a < 1$); details
can be found in the Appendix. This procedure is similar to the one
proposed in Ref. \cite{Wang} and later adopted in Refs. \cite{Samsonov,
Avinash}. One obtains
\[
\omega_{L, 0}^2 = \frac{2 Q^2}{M \lambda_D^3} \,
\frac{1}{\kappa^3}  \, \times \qquad \qquad \qquad \qquad \qquad \, \]
\begin{equation}
\biggl[ e^{-\kappa/2}\,\frac{\kappa}{2}\,
csch\biggl(\frac{\kappa}{2}\biggr) \,+ \,\frac{\kappa^2}{8}\,
csch^2 \biggl(\frac{\kappa}{2}\biggr) \,- \,\ln ( 1 - e^{-\kappa} )
 \biggr]
\, ,
\label{Ninfinite-omega0}
\end{equation}
for the characteristic oscillation frequency $\omega_{L, 0} =
v_0/(\kappa \lambda_D)$; the result for $v_0$ is obvious;
csch$x = 1/\sinh x$. A numerical
investigation shows that the numerical value of the frequency in
the region near $r_0 \approx \lambda_D$ (i.e. $\kappa \approx 1$)
is thus increased by a factor of $1.5$ or higher, roughly, compared to
the NNI expression above (see Fig. \ref{figure1}), and so does the
characteristic second-order dispersion velocity $v_0^2 =
\omega_{L, 0}\,r_0 = \omega_{L, 0}\,\lambda_D\, \kappa$
(see Fig. \ref{figure2}). A similar effect is witnessed for the
characteristic velocity $v_1$, related to the
fourth-order dispersion
\begin{eqnarray}
v_{1}^2 = \frac{Q^2}{M \lambda_D} \, \frac{1}{96 \kappa}
csch^4\biggl(\frac{\kappa}{2}\biggr) \, \times \qquad \qquad \qquad
\qquad \qquad \nonumber \\
\biggl[ ( {\kappa}^{2}+ 2
)\, cosh\kappa  \,+ \,2 \,( \kappa^2 -1  \,+ \,\kappa \, sinh
\kappa )
 \biggr]
\, \, ,
\label{Ninfinite-v12}
\end{eqnarray}
(see Fig. \ref{figure3}) and for the nonlinearity coefficients
\begin{eqnarray}
p_{0} = \frac{Q^2}{M \lambda_D} \,
 \biggl\{ \frac{1}{(e^\kappa
- 1)^3} \, \biggl[ 6 \, + \, e^\kappa ( {\kappa}^{2} - 3 \kappa -
12 )\, \qquad \nonumber \\
+ \,e^{2 \kappa} \,( \kappa^2 \,+ \,3 \kappa \, +6 )
 \biggr]
\,- \frac{6}{\kappa} \,\ln \bigl( 1+ \sinh \kappa - \cosh \kappa
\bigr) \biggr\} \, ,
\label{Ninfinite-p0}
\end{eqnarray}
and
\begin{eqnarray}
q_{0} = \frac{Q^2}{M \lambda_D} \,
 \biggl\{ \frac{1}{(e^\kappa
- 1)^4} \, \biggl[ - 12 \, \qquad \qquad\qquad\qquad \nonumber
\\
+ \, e^{2 \kappa} [( {\kappa}^{3} + 12
\kappa + 48 )\cosh\kappa \nonumber
\\  + \,2 ( \kappa^3 - 6 \kappa - 18) + 2 (\kappa^2-6) \sinh\kappa
]
 \biggr] \nonumber
\\
\, - \frac{12}{\kappa} \,\ln \bigl( 1+ \sinh \kappa - \cosh \kappa
\bigr) \biggr\} \, .
\label{Ninfinite-q0}
\end{eqnarray}
Upon simple inspection of Figs. \ref{figure4} and \ref{figure5},
one deduces that $q_0$ takes practically double the value of $p_0$
everywhere, and thus draws the conclusion that $q_0$ should rather
\emph{not} be omitted in Eq. (\ref{eqmotion-gen-continuum}) (cf. e.g.
Refs. \cite{Melandso, Samsonov, Avinash, PKS2003}), for the case of
the Debye potential.

\section{Linear oscillations}

Let us first consider the
linear regime in longitudinal grain oscillations.
For the sake of rigor, one may revert to the discrete formula
(\ref{eqmotion1}) and consider its linearized form by simply
neglecting the two last (double) sums therein. Inserting the ansatz
$\delta x_n \sim \exp i(n k r_0 - \omega t)$, where $\omega$ is the
phonon frequency and
$k = 2 \pi/\lambda$ (respectively, $\lambda$) is the wavenumber
(wavelength), one immediately
obtains the general dispersion relation
\begin{eqnarray}
\omega \, ( \omega + i \, \nu ) \, = \,
\frac{4 Q}{M}\, \sum_{l = 1}^N\, \phi''(l r_0)\,
\sin^2\frac{l \, k r_0}{2}
\, \nonumber \\
= \,
\frac{4 Q}{M}\, \sum_{l = 1}^N\,
\phi''(l \kappa \lambda_D)\, \sin^2\frac{l \, \kappa
\,(k \lambda_D)}{2} \, .
\label{dispersion-discrete}
\end{eqnarray}
One may readily verify that the standard 1D acoustic wave dispersion
relation $\omega \approx v_0\, k$ is obtained in
the small $k$ (long wavelength) limit: check by setting
$\sin({l \, k r_0}/{2}) \approx {l k r_0}/{2}$
(and recalling the general definition of
$v_0$ above).
Of course, taking this limit
simply amounts to linearizing the continuum equation
of motion derived above (and keeping the lowest contribution in $k$).
As pointed out before (see e.g. Ref. \cite{Wang}),
one thus recovers the dust-acoustic wave dispersion relation obtained
in the strong-coupling dusty plasma regime
(upon defining the density $n_d$ as $\sim r_0^{-3}$, which may
nevertheless appear somehow heuristic in this 1D model).

\begin{figure}[htb]
 \centering
 \resizebox{3in}{!}{
 \includegraphics[]{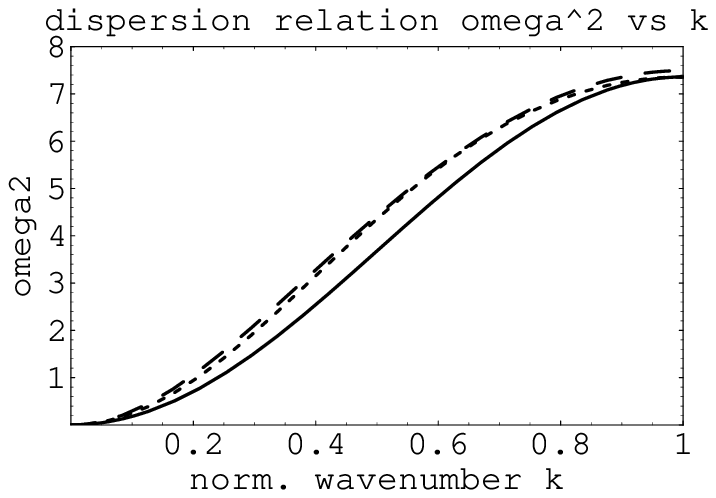}}
\caption{} \label{figure6}
\end{figure}

Notice that the form of the dispersion relation, in principle, depends
on the value of $N$. However, in the case of the Debye interactions, i.e.
explicitly substituting $d^2 \phi_D(x)/dx^2$ into
(\ref{dispersion-discrete}), one obtains
\begin{equation}
\omega \, ( \omega + i \, \nu ) \, = \,
\frac{4 Q^2}{M \lambda_D^3}\,
\sum_{l = 1}^N\,  e^{-l \kappa}\,  \frac{2 + 2 \kappa+
{(l \kappa)^2}}{(l \kappa)^3}\,
\sin^2\frac{l \, k r_0}{2}
 \, .
\label{dispersion-Debye}
\end{equation}
A numerical investigation, e.g. for $\kappa = 1$
(see Fig. \ref{figure6}), suggests that the dispersion curve quickly
sums up to a limit curve, even for not so high values of $N$ (practically
for $N = 2$ already). The values of the frequency reduce with increasing
$\kappa$, as suggested by the exponential term.
We see that the dispersion
curve possesses a maximum at $k = \pi/r_0 = \pi/(\kappa \lambda_D)$
 for any
value of $\kappa$ and $N$.

The dispersion curves of dust-lattice waves have been investigated
by both experiments (see e.g. Refs. \cite{Homann1998, Nunomuradisp})
and \textit{ab initio}
numerical simulations \cite{sim}.
In should nevertheless be acknowledged that the results of these studies
do not absolutely confirm the dispersion curves obtained above,
which suggests, as poined out in Ref. \cite{Homann1998},
that one-dimensional crystal models may be inappropriate for real dust
crystals.

\section{The Korteweg -- De Vries (KdV) Equation}

In order to take into account weak nonlinearities,
a procedure which is often adopted at a first step
consists in keeping only the
first nonlinear contribution in Eq. (\ref{eqmotion-gen-continuum})
(by cancelling the last term in the rhs, i.e. setting $q_0 = 0$) and
then considering excitations moving at a velocity close to the characteristic
velocity $v_0$.
A Galilean variable transformation, viz.
\begin{equation}
x \rightarrow \zeta = x - v_0\, t \, , \qquad t \rightarrow \tau = t
\, , \qquad w \, = \,
u_\zeta \, , \label{transfo1}
\end{equation}
then provides the {\sc{Korteweg -- De Vries}} (KdV) Equation
\begin{equation}
w_\tau \,- \, s\, a\, w \, w_\zeta + \, b\, w_{\zeta\zeta\zeta}\, = \, 0 \,
,
\label{KdV}
\end{equation}
where a term $u_{\tau\tau}$ was assumed of
higher-order and thus neglected.
The coefficients are
\begin{equation}
a\,  = \, \frac{|p_0|}{2 v_0} \, , \qquad b\,  = \,
\frac{v_1^2 r_0^2}{2 v_0}
\,
, \qquad s = sgn \,p_0 = p_0/|p_0| \, .
\label{coeffsKdV}
\end{equation}
We have introduced the parameter $s$ ($= +1/-1$),
denoting the sign of $p_0$, which may change the form of the solutions
(see below); as
discussed above, it is equal to $s = +1$ for the Debye-type
interactions.
 It should be noted that
this procedure is identical to the one initially adapted for
dust-lattice-waves in \cite{Melandso} and then followed in
Ref. \cite{Samsonov, Avinash, PKS2003} (for $s = + 1$) as may readily be
checked, yet the new aspect here lies in the generalized
definitions of the physical parameters above. Also notice
that positive-oriented ($\sim \hat x$) propagation was considered;
adopting the above procedure in backward ($\sim - \hat x$)
propagation is trivial, yet it should be carried out
by re-iterating the analytical procedure and
\emph{not} by plainly considering $v \rightarrow -v$: the KdV equation is
\emph{not symmetric} with
respect to this transformation (also see that the
velocity $v$ appears under a square root in the formulae).

\begin{figure}[htb]
 \centering
 \resizebox{3in}{!}{
 \includegraphics[]{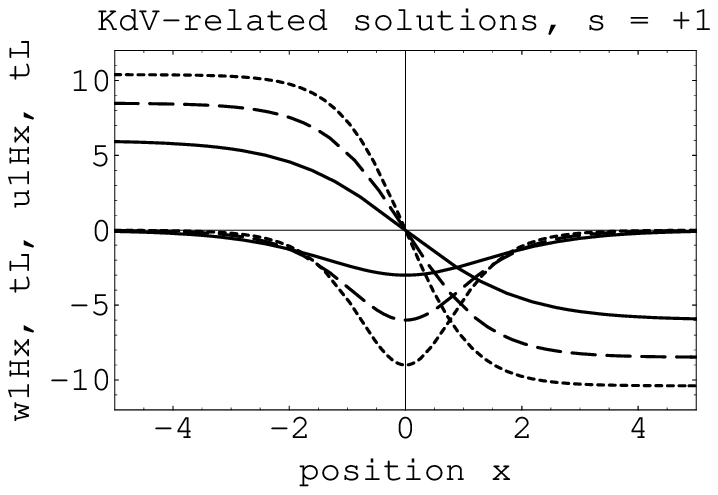}} \\
\vskip 2 cm
\resizebox{3in}{!}{
 \includegraphics[]{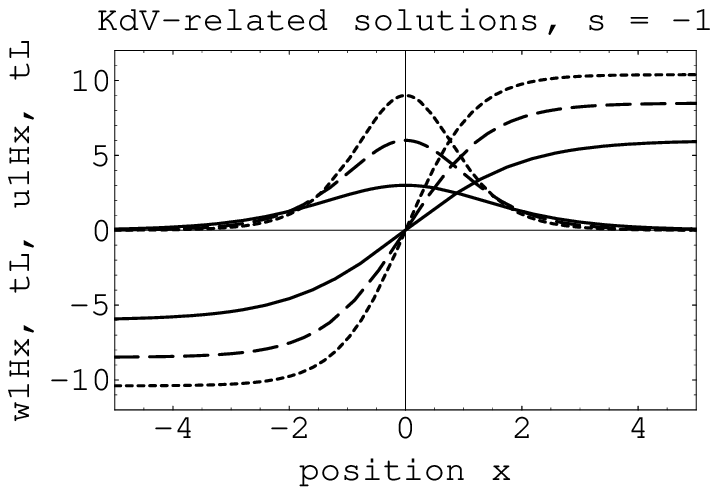}}
\caption{} \label{figure7}
\end{figure}

As a mathematical entity,
the KdV Equation has been extensively studied \cite{Drazin, Remo,
Dodd, Sagdeev, Davydov-comment, Newell, Karpman, Lonngren, Whitham}, so only
necessary details will be summarized here. It is known to
possess a rich variety of solutions, including periodic
(non-harmonic) solutions (cnoidal waves, involving elliptic
integrals) \cite{Sagdeev}. For vanishing boundary conditions, Eq.
(\ref{KdV}) can be shown (see e.g. in Refs. \cite{Drazin, Remo})
to possess one- or more ($N-$) soliton localized solutions
$w_N({\zeta, \tau})$ which bear all the well-known soliton
properties: namely, they propagate at a constant profile, thanks to
an exact balance between dispersive and nonlinear effects, and
survive collisions between one another. The simplest (one-)
soliton solution has the pulsed-shaped form
\begin{equation}
w_1({\zeta, \tau})\, = \,-s \, w_{1, m} sech^2 \biggl[
(\zeta - v \tau - \zeta_0)/L_0 \biggr] \, ,
\label{KdVw1sol}
\end{equation}
where $x_0$ is an arbitrary constant, denoting the initial soliton
position, and $v$ is the velocity of propagation;
in principle, $v$ may take any real value even though its
 range may be physically limited, as in our case, where $v$ has been
 assumed close to $v_0$; this constraint will be relaxed below.
A qualitative result to be retained from the soliton solution in
(\ref{KdVw1sol})
is the velocity
dependence of both soliton amplitude $w_{1, m}$ and width $L_0$, viz.
\[w_{1, m} = 3 v/a = 6 v v_0/|p_0|\, , \]
\[L_0
= {(4 b/v)}^{1/2} = [2 v_1^2 r_0^2/(v v_0)]^{1/2} \, .\]
We see that $w_{1, m} L_0^2 =
constant$, implying that narrower/wider solitons are taller/shorter
and propagate faster/slower. These qualitative aspects of
dust-lattice solitons have recently been confirmed by dust-crystal
experiments \cite{Samsonov}.
Notice that the solutions of
(\ref{KdV}) satisfy an infinite set of conservation laws \cite{Drazin,
Karpman}; in particular, the solitons $w_N$ carry a constant
`mass' $M \sim \, \int w d\zeta$ (which is negative for a negative
pulse), `momentum' $P \sim \, \int w^2 d\zeta$,
 `energy' $P \sim \, \int (w_x^2/2 + u^3)\, d\zeta$,
and so forth (integration is understood over the entire $x-$ axis)
\cite{Davydov-comment, Drazin}. See that the forementioned
amplitude--width dependence of the 1-soliton solution
(\ref{KdVw1sol}) is heuristically deduced from the soliton `mass'
conservation law (implying conservation of the surface under the
bell-shaped curved in Fig. \ref{figure7}): taller excitations have to be
thinner and vice versa.

Inverting back to our initial reference frame,
one obtains, for the spatial
displacement variable $u(x, t)$, the kink/antikink (for $s = -1/+1$)
solitary wave form
\begin{equation}
u_1(x, t)\, = \,- s\,
u_{1, m}
\,\tanh \, \biggl[ (x - v t - x_0)/L_1 \biggr] \, ,
\label{KdVu1kink}
\end{equation}
which represents a localized region of compression/rarefaction
(for $s = +1/-1$), propagating to the positive direction of the
$x-$axis (see Fig. \ref{figure7}). The amplitude $u_{1, m}$ and the
width $L_1$ of this shock excitation are
\[u_{1, m} \, = \,
\frac{6 v_1 r_0}{|p_0|}\,\bigl[ 2 \, v_0 \, (v - v_0) \bigr]^{1/2}
\, , \]
\[ L_1 \, = \,
r_0  \biggl[ \frac{2 \,v_1^2}{v_0\,(v - v_0)} \biggr]^{1/2}
\, \,=
\frac{12 v_1^2 r_0^2}{|p_0|} \, \frac{1}{A_m} \, ,
\]
imposing `supersonic' propagation ($v > v_0$) for stability,
in agreement with experimental results in dust
crystals \cite{Samsonov}.
Notice that faster solitons will be narrower,
and thus more probable to `feel' the lattice discreteness,
contrary to the continuum assumption above; therefore, one
may impose the phenomenological criterion: $L \ll r_0$,
amounting to the condition $v/v_0 \gg 1 + 2 \, v_1^2/v_0^2$
[$\approx 1.17$ for the Debye NNI case; see (\ref{om0NNI}) above],
in order for the above (continuum) solution to be sustained in the (discrete) chain.
Nevertheless, supersonic wave stable propagation has been
numerically verified at a wide range of velocity values
in atomic chains \cite{Pnevmatikos-thesis, Pnevmatikos1986}, where
Eqs. (\ref{eqmotion-gen-continuum}) and (\ref{KdV}) arise via
a procedure similar to the one outlined above; also see
Ref. \cite{Hao} for a recent experiment in crystalline solids.
Finally, note that $v_0$ in real DP crystals
bears values as low as a few tens of mm/sec \cite{Samsonov,
Nosenko}.

Remarkably, Eq. (\ref{KdV}) is exactly solved
 via the Inverse Scattering Transform \cite{ISM, Drazin, Newell}, for any given
initial condition $u({\zeta, 0})$, which is generally seen to break-up into
a number of (say $N$, depending on $u({\zeta, 0})$ \cite{ISM}) solitons plus
a tail of background oscillations.
These considerations, including, in particular,
the two-soliton solution $w_2$ of Eq. (\ref{KdV}),
which represents two distinct humps moving at different velocities and
colliding during propagation without changing shape,
have been postulated to be of relevance in the interpretation of
recent dusty plasma discharge
experiments \cite{Samsonov, Avinash}.

The wide reputation of the KdV Equation (\ref{KdV}) is mostly due
to the exhaustive knowledge of its analytical properties
\cite{Drazin, Dodd, Newell, Remo, Davydov-comment, Sagdeev,
Karpman}, in addition to its omni-presence in a variety of
physical contexts, not excluding the physics of ordinary (ideal,
i.e. electron--ion) plasmas \cite{Karpman, Debnath} and, more recently,
dusty plasmas \cite{PSbook, PKS2003}. However, in the above
dusty-plasma-crystal context, it has been derived under specific
assumptions (low discreteness and low nonlinearity
effects; also, a propagation velocity $v \approx v_0$) which
may be questionable, in a real DP crystal. Even if the first one
is virtually impossible to cope with, analytically, the latter
ones may be somehow relaxed via a different approach, to be
outlined below.

\section{Higher-order Korteweg--DeVries (EKdV) Equations}

In order to derive a KdV equation from the continuum equation of motion
(\ref{eqmotion-gen-continuum}),
we have neglected the coefficient $q$, which is related to the
cubic nonlinearity of the interaction potential.
Nevertheless, a simple numerical investigation shows that this term is {\em{not}}
small, and may, in certain cases, even dominate over the quadratic term $p$, as
in the Debye potential case (see the discussion above).
Therefore, one is tempted to find out how the dynamics is modified if this term
is taken into account.

\subsection{The Extended Korteweg -- de Vries (EKdV) Equation}

Repeating the procedure which led to
Eq. (\ref{KdV}),
in the previous section, yet now keeping
the fourth order derivative coefficient $q \ne 0$ in
Eq. (\ref{eqmotion-gen-continuum}), one obtains the
EKdV Equation
\begin{equation}
w_\tau \,- \, s\, a\, w \, w_\zeta
+ \,  \hat a\, w^2 \, w_\zeta
+ \, b\, w_{\zeta\zeta\zeta}\,
 = \, 0 \, ,
\label{EKdV}
\end{equation}
where all coefficients are given in (\ref{coeffsKdV}) except $\hat
a = q_0/(2 v_0)$; recall that $a$, $b$ are positive by definition.
We shall see below that $p_0, q_0 > 0$ for Debye
interactions (yet not necessarily, in general), so that the
nonlinearity coefficients, i.e. $- s a$ (for $s = +1$)
and $\hat a$, bear negative and positive (respectively)
values in this (Yukawa crystal) case.

\begin{figure}[htb]
 \centering
 \resizebox{3in}{!}{
 \includegraphics[]{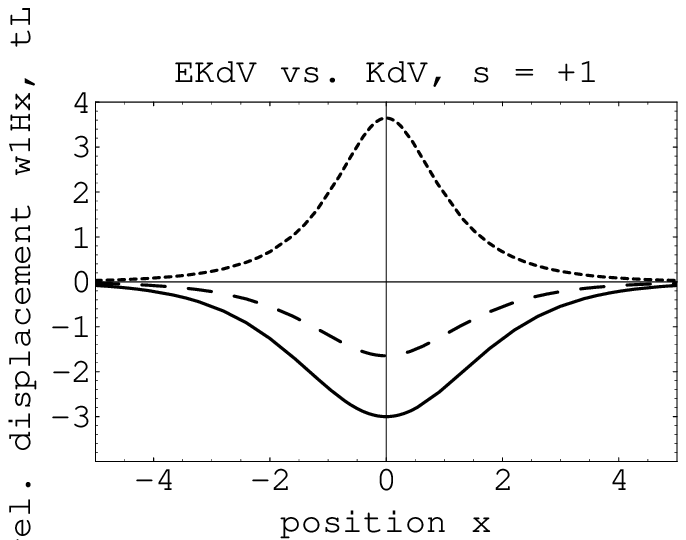}} \\
\vskip 2 cm \resizebox{3in}{!}{
 \includegraphics[]{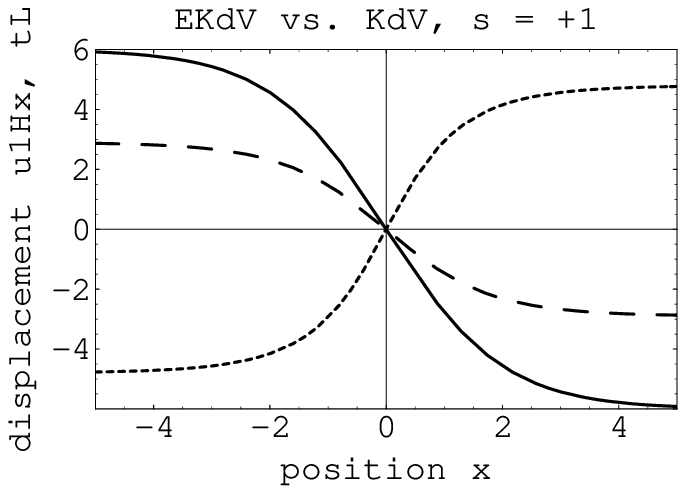}}
\caption{} \label{figure8}
\end{figure}

The EKdV Eq. (\ref{EKdV}) was thoroughly studied in a classical
series of papers by Wadati \cite{Wadati}, who derived it for
nonlinear lattices, then obtained its travelling-wave and,
separately, periodic (cnoidal wave) solutions and, finally,
exhaustively studied its mathematical properties. Both
compressional and rarefactive solitons (say, $w_{2, \pm}$, to be
distinguished from the KdV solution $w_{1}$) were found to solve
Eq. (\ref{EKdV}) (for either signs of $s$); adapted to our
notation here \cite{commentWadati1}, they are of the form
\begin{eqnarray}
w_{2}^{(1)}(\zeta_, \tau) = - s \, {v}/\bigl\{ C \cosh^2 [ (\zeta -
v \tau - \zeta_0)/L_0] \qquad
\nonumber \\
\qquad
+ \, D \sinh^2 [(\zeta - v \tau -
\zeta_0)/L_0] \bigr\} \, , \label{EKdVsol1}
\end{eqnarray}
and
\begin{eqnarray}
w_{2}^{(2)}(\zeta_, \tau) = + s \, {v}/ \bigl\{
D \cosh^2 [(\zeta - v
\tau - \zeta_0)/L_0] \qquad
\nonumber \\
\qquad + \, C \sinh^2 [(\zeta - v \tau - \zeta_0)/L_0] \bigl\}
\, , \label{EKdVsol2}
\end{eqnarray}
where
\begin{eqnarray}
C \,&=& \,\frac{a}{6} \, \biggl( \sqrt{1 + \frac{6 \hat a v}{a^2}} +
1 \biggr) \nonumber \\
&= &\, \frac{1}{12 v_0} \, \biggl( \sqrt{p_0^2 + 1 2 q_0
v_0 v}  + |p_0| \biggr)
\, , \nonumber \\
D \,&=& \,\frac{a}{6} \, \biggl( \sqrt{1 + \frac{6 \hat a v}{a^2}} -
1 \biggr) \nonumber \\
&=& \, \frac{1}{12 v_0} \, \biggl( \sqrt{p_0^2 + 1 2 q_0
v_0 v}  - |p_0| \biggr)\,,
\end{eqnarray}
the width $L_0$ was defined above, and $v > 0$ is the propagation
velocity. For $s=+1/-1$, the first expression represents a
propagating localized compression/rarefaction, while the second
denotes a (larger, see comment below; cf. Fig. \ref{figure8})
rarefaction/compression, respectively. Notice that, for $q_0 \sim
\hat a = 0$, the first expression recovers the KdV result obtained
previously (since $v/C$ then recovers the KdV soliton width $w_{1,
m}$), while the second results in a divergent (physically
unacceptable) solution \cite{Wadati}. Following Wadati, we may
re-arrange (\ref{EKdVsol1}) and (\ref{EKdVsol2}) as
\begin{eqnarray}
w_{2}^{(j)}(\zeta, \tau) = - s \, \epsilon_{j} \frac{2 \sqrt{6
b}}{\sqrt{\hat a}} \, \times \qquad \qquad \qquad \qquad \qquad
\nonumber \\
\qquad \frac{\partial }{\partial \zeta} \biggl\{
\tan^{-1} \biggl[ \tilde W_2^{(j)} \, \tanh \biggl( \frac{\zeta -
v \tau - \zeta_0}{L_0} \biggr) \biggr] \biggr\} \label{EKdVsol12}
\, ,
\end{eqnarray}
provided that $\hat a \sim q_0 \ne 0$. Here
\begin{equation}
\tilde W_2^{(j)}\, = \, \biggl(
\frac{\sqrt{1 + \frac{6 \hat a v}{a^2}} - \epsilon_{j}}
{\sqrt{1 + \frac{6 \hat a v}{a^2}} + \epsilon_{j}} \
\biggr)^{1/2} \, . \label{tildeW2}
\end{equation}
Furthermore, $v > 0$ is the propagation velocity in the frame $\{ \zeta,
\tau \}$ and $j =1$ ($2$) recovers $w_{2}^{(1)}$ ($w_{2}^{(2)}$) above,
so that $\epsilon_1$ ($\epsilon_2$) is equal to $+1$ ($-1$),
representing rarefactive (compressive) solutions, for $s=+1$ -- e.g.
the Debye case -- and vice versa for $s=-1$ (which recovers Wadati's
notation). The pulse width now depends on both $L_0$ (defined as previously)
and $\tilde W_2^{(j)}$.
The pulse value for $s = -1$ \cite{Wadati} satisfies:
\[
w_{-} \equiv - (\sqrt{a^2 + {6 \hat a v}} + a) \, < \, w_{2}^{(1)} < 0 \, < \,
w_{2}^{(2)} \, \]
\[
\qquad \qquad\qquad < \, (\sqrt{a^2 + {6 \hat a v}} - a) \,  \equiv w_{+}
\, ,
\]
(for $s = +1$, one should permute the superscripts 1 and 2); since
$|w_{-}| > |w_{+}|$, one expects, for $s = -1$, a small
rarefactive and a large compressive pulse; the opposite holds for
$s = +1$, e.g. in a Debye crystal case: see Fig. \ref{figure8}.

Inverting to the lattice displacement coordinate $u \sim \int w
d\zeta$, expressed in the original coordinates $\{ x, t \}$, we
obtain
\begin{eqnarray}
u_{2}^{(j)}(x, t) = - s \, \epsilon_{j} \, 2 \, \sqrt{\frac{6
v_1^2}{q_0}} \, r_0 \, \times \qquad \qquad\qquad\qquad\nonumber \\
\tan^{-1} \biggl[ W_2^{(j)} \, \tanh
\biggl( \frac{x - v t - x_0}{L_1} \biggr) \biggr] \, ,
\label{EKdVu12}
\end{eqnarray}
where
\begin{equation}
W_2^{(j)}\, = \, \biggl( \frac{\sqrt{p_0^2 + \, 12 q_0 \, v_0 \,
(v - v_0)} - \epsilon_j \,|p_0|} {\sqrt{p_0^2 + \, 12 q_0 \, v_0
\, (v - v_0)} + \epsilon_j \,|p_0|} \biggr)^{1/2} \, , \label{W2}
\end{equation}
and $j = 1, 2$.
As expected, for any given $s$ ($= \pm 1$), the two different
kink/antikink solutions obtained for different $j$ ($=1$ or $2$) are
not symmetric; cf. Figure \ref{figure8}. Notice that the maximum value now
also depends on $W_2^{(j)}$ ($j = 1, 2$).

In conclusion, the Extended KdV equation provides a more complete
description of the nonlinear dynamics of the lattice, compared to
the KdV equation. In particular, the EKdV compressive
(rarefactive) pulse soliton obtained for $s = +1$ ($s = -1$), i.e.
$p_0 > 0$ ($p_0 < 0$) is slightly smaller than its KdV counterpart
(see Fig. \ref{figure8}), but the EKdV also predicts the
possibility for a rarefactive (compressive) soliton, in either
case, to form and propagate in the same lattice. In the particular
case of Debye crystals, the net new result to be retained is the
prediction of the existence of a rarefactive new excitation, in
addition to the rarefactive one, observed in experiments.
Nevertheless, theoretical studies on molecular chains seem to
suggest that the additional shock-like localized mode predicted by
the EKdV equation will not be as stable as its (KdV--related)
counterpart. This prediction should, therefore, be confirmed
numerically (and experimentally) before being taken for granted.

\subsection{The Modified Korteweg -- de Vries (MKdV) Equation}

Note, for the sake of rigor, that upon setting $p_0 \sim a = 0$ in
Eq.(\ref{EKdV}) above, one obtains a modified KdV
(MKdV) equation (with only a cubic nonlinearity term). The MKdV
equation shares all the qualitative properties of the KDV Eq. and is,
in fact, related to it via a Miura transformation \cite{Drazin}.
It has two (both negative and positive, for each value of $s$)
pulse soliton solutions which follow immediately from the
preceding solutions (\ref{EKdVsol1}) and (\ref{EKdVsol2}) of the
EKdV equation, upon setting $p_0 = 0$. The remarkable additional aspect of the
MKdV equation is that it also bears slowly oscillating solutions, named
{\em{breathers}}, obtained just as rigorously via the inverse
scattering method \cite{Lamb, Flytzanis1985}. These solutions
(whose wavelength is comparable to their localized width, hence
the `breathing' impression and the name) share the remarkable
properties of solitons; in particular, they are seen to survive
collisions between themselves and with pulse solitons
\cite{Flytzanis1985}. Their analytic form can be readily found in
Refs. \cite{Flytzanis1985} (see \S 4.1 therein) and \cite{Lamb}
and will not be reproduced here, since their condition of
existence, namely $p_0 \ll q_0$ (cf. Ref. \cite{Flytzanis1985})
is rather {\em{not}} satisfied in the case of
Debye-interacting dust grains. Note however that breather-like
excitations may exist in a DP crystal, as one may see via a
different (perturbative) analysis of the nonlinear modulation
of the amplitude of longitudinal lattice waves. This is considered in separate
work \cite{IKPSLDLWNLS}.

\section{The Boussinesq (Bq) and Generalized Boussinesq (GBq) Equations}

Remember that the KdV--type equations in the preceding Section
were obtained from the equation of
motion (\ref{eqmotion-gen-continuum}) in an approximative manner,
assuming near--sonic propagation and neglecting high--order time derivatives.
Those results are therefore expected to hold for velocity values
only slightly above $v_0$.
We shall now see how these assumptions can be relaxed by directly
relying on the initial (nonlinear) equation .

Let us consider Eq. (\ref{eqmotion-gen-continuum}) again (for $\nu = 0$).
Upon setting $p_0 = - 2 p$, $q_0 = 3 q$, $v_1^2 r_0^2 = h > 0$,
and integrating
once, with respect to $x$, one
exactly obtains, for $w = u_x$, the {\sc{generalized Boussinesq}} (GBq)
Equation
\begin{equation}
\ddot{w}  \,- v_0^2 \, w_{xx}\,= h \, w_{xxxx}\,
+ \, p \, (w^2)_{xx} \, + \, q \, (w^3)_{xx}
\label{GB}
\end{equation}
which, neglecting the cubic nonlinearity coefficient
$q$ [viz. $q_0 = 0$ in
(\ref{eqmotion-gen-continuum})], reduces to the well-known
Boussinesq (Bq) equation, widely studied e.g. in solid
chains; see e.g. \cite{Flytzanis1985, Pnevmatikos-thesis,
Pnevmatikos1986}. It possesses well-known localized solutions, whose
derivation is straightforward and need not be reproduced here. The
exact expressions obtained from (\ref{GB})
for the relative displacement $w(x, t)$ and
the longitudinal displacement $u(x, t)$
are exhaustively presented and discussed in Refs. \cite{Flytzanis1985,
Pnevmatikos1986}. The analytic kink/antikink-type localized solutions
for $u(x, t)$ read
\begin{eqnarray}
u_{3}(x, t)\, = \, \mp 2 \, sgn(h)\, \biggl( \frac{6 h}{q_0}
\biggr)^{1/2} \, \times \qquad  \qquad  \qquad \nonumber \\
\tan^{-1}\, \biggl[W_3 \, \tanh \, \frac{x
- v t - x_0}{L_3} \biggr]\, . \label{GBqkink}
\end{eqnarray}
Here the soliton velocity is $v$, while the soliton width depends
on both $W_3$ and $L_3$, which are
\begin{eqnarray}
W_3\, = \, \biggl\{ \frac{[p_0^2 + 6 q_0 (v^2 - v_0^2)]^{^1/2} \mp
|p_0|} {[p_0^2 + 6 q_0 (v^2 - v_0^2)]^{^1/2} \pm |p_0|} \biggr\}^{1/2}
\, , \nonumber \\
 L_3 \,= \, 2\, \biggl( \frac{h} {v^2 - v_0^2}
\biggr)^{1/2} \, . \qquad \qquad \qquad \qquad \label{GBqkink-defs}
\end{eqnarray}

Recall that, for Debye interactions, $h, q > 0$ and $p = - p_0 <
0$ (see above), prescribing the `supersonic' ($v > v_0$)
propagation of the solutions; the same was true of the KdV
solitons obtained above. Notice, however, that the expressions
obtained here for the longitudinal displacement $u$ represent
both rarefactive and compressive lattice excitations even for
$p_0 > 0$ (see Table I in \cite{Pnevmatikos1986}, for $p = - p_0 <
0$); remember that this feature was absent in the KdV equation
(\ref{KdV}), where $p_0 > 0$ i.e. $s = + 1$ always led to a
compressive solution in (\ref{KdVu1kink}). In fact, this is
also true of the
Bq equation, which is
obtained for $q_0 = 0$, i.e. neglecting the last term in the
continuum equation of motion (\ref{eqmotion-gen-continuum}). The
exact solution then reads
\begin{eqnarray}
u_{Bq}(x, t)\, = \, - sgn(h)\, sgn(p_0)\,
\frac{6 [h (v^2 - v_0^2)]^{1/2}}{|p_0|}
 \, \times \qquad \nonumber \\
\tanh \, \frac{x - v t - x_0}{L_1} \, , \qquad
\label{Bqkink}
\end{eqnarray}
which, for positive $h$ and $p_0$, prescribes only compressive
supersonic kinks, pretty much like the solution $u_1$ derived  from
the KdV theory above
(cf. Table I in \cite{Pnevmatikos1986}, for $h_0
> 0$, $p < 0$ and $q = 0$).

Closing this section, one may wish
to compare the GBq and Bq solutions (\ref{GBqkink}) and (\ref{Bqkink}),
to the homologous EKdV-- and KdV--related
solutions (\ref{EKdVu12}) and (\ref{KdVu1kink}), respectively,
obtained previously:
one may readily check that the former ones tend to the latter two as $v$
tends to $v_0$ [to see this, one may set \
$v^2 - v_0^2 = (v + v_0)(v^2 - v_0) \approx 2 v_0 (v^2 - v_0)$;
recall that $h = v_1^2 r_0^2$]. Nevertheless, this
velocity range restriction is relaxed in the Boussinesq--related
description.

\section{Excitations in real DP crystals}

It is now quite tempting to observe and compare the predictions
furnished by the above nonlinear models in a DP crystal in terms of
excitation features, e.g. dimensions and form.
For instance, one may substitute the expressions for the
model's physical parameters
(i.e. $\omega_0$, $v_0$, $v_1$, $p_0$ and $q_0$)
derived in \S \ref{DebyeDPC} into
the definitions in the latter three sections, in order to derive a final
form for localized excitations in a real DP crystal, in terms of the
propagation velocity $v$, the lattice parameter $\kappa$ and, generally,
the sign $s$ ($= +1$ for Debye interactions).
The interest in this procedure is evident, since one may seek feedback
(e.g. parameter values, excitation behaviour) from experiments and then
investigate the validity of the above models by adjusting them to real
DP crystal values.

The final expressions for $u_j(x, t)$
[$j = 1, 2, 3$, cf. (\ref{KdVu1kink}), (\ref{EKdVu12}) and
(\ref{GBqkink}), respectively]
are somewhat lengthy and need not be reported here (since they are
straightforward to derive). We
may nevertheless summarize some interesting numerical results.

\begin{figure}[htb]
 \centering
 \resizebox{3in}{!}{
 \includegraphics[]{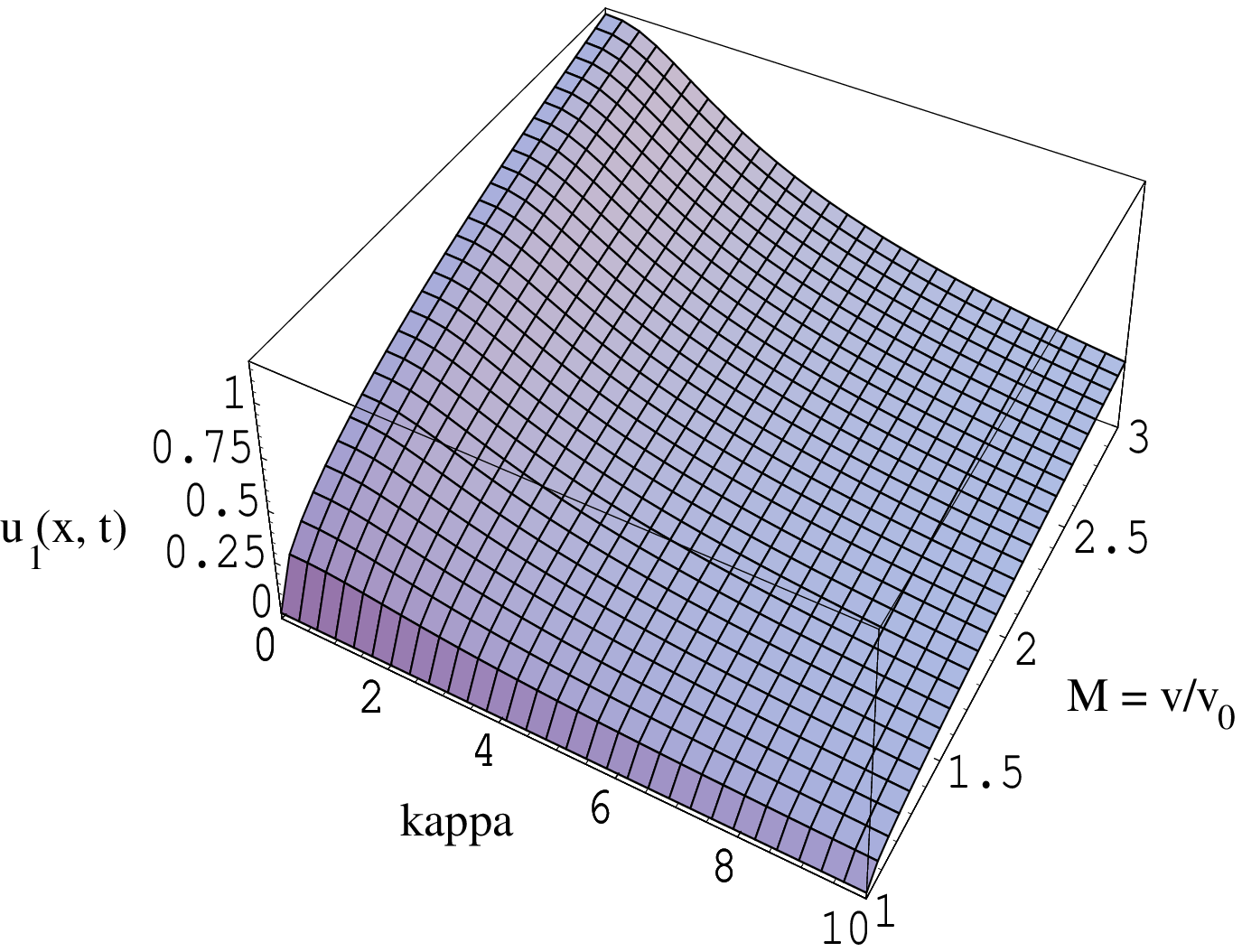}} \\
\vskip 2 cm \resizebox{3in}{!}{
 \includegraphics[]{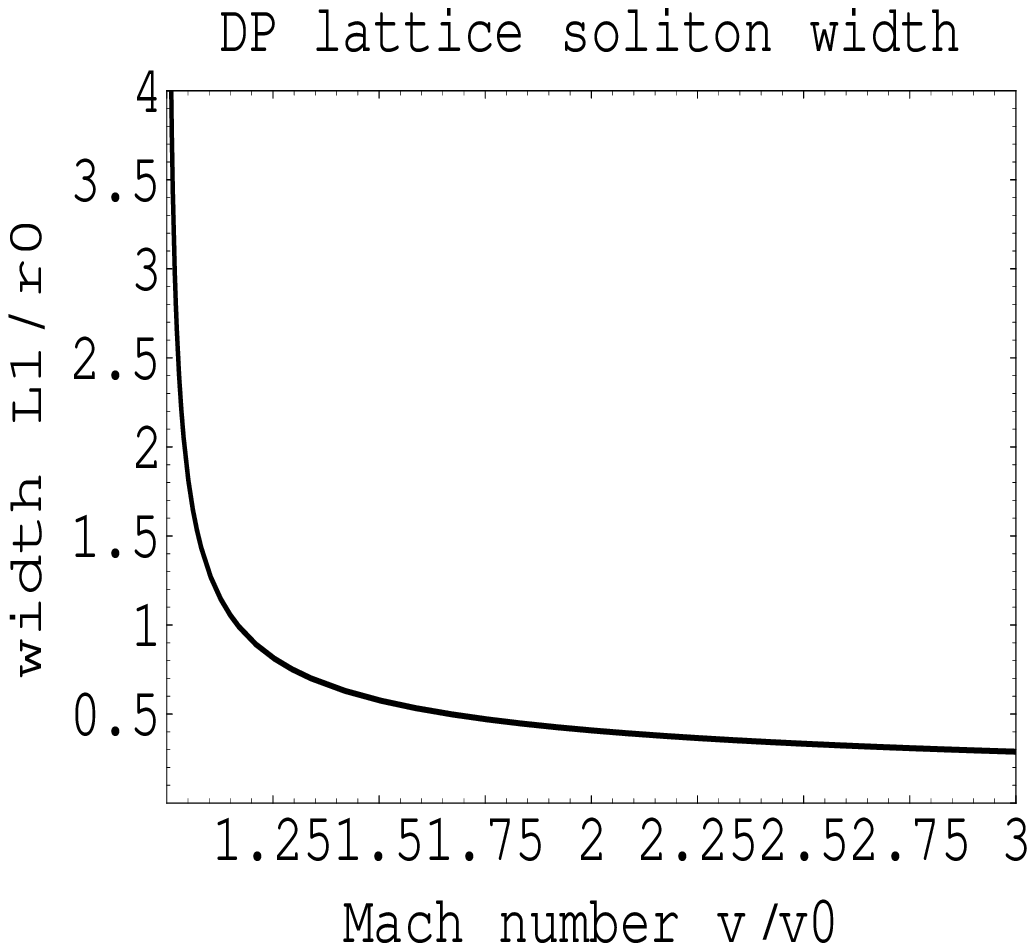}}
\caption{} \label{figure9}
\end{figure}

The soliton width $L_1$, as defined in (\ref{KdVu1kink}), now becomes
$L_1 = v_0^2/p_0 u_{1, m}$: we see that the product of the displacement $u$
kink's width and maximum value remains
constant (regulated by the cubic interaction potential nonlinearity),
viz. $u_{1, m} \, L_1 \, = 1/p_0$, unlike the KdV pulse
soliton for the relative displacement $w = u_x$
which is characterized by
$w_{1, m} \, L_1^2 \, = cst.$.
Both the kink maximum value $u_{1, m}$ and width $L_1$
depend on the velocity $v$; as a matter of fact, faster kink
excitations will be taller and narrower - see Fig. \ref{figure9} -
since now
\begin{equation}
\frac{u_{1, m}}{r_0} \, = \, \frac{v_0^2 \sqrt{6}}{p_0} \sqrt{M - 1} \, ,
\qquad
\frac{L_1}{r_0}  \, = \, \frac{1}{\sqrt{6 (M - 1)}} \, .
\end{equation}
Recall that the Mach number $M = v/v_0$ is always larger than unity.
Furthermore, the magnitude of the excitation seems to
decrease with $\kappa$; see Fig.\ref{figure9}a: nevertheless,
very high values (near $u/r_0 = 1$)
observed for low $\kappa$ and high $v$ are rather not to be trusted,
since they contradict the continuum approximation $u \ll r_0$.

Finally, one may compare the solutions obtained from the above theories,
for a typical value of $\kappa$, say $1.25$, according with real
experimental values. The KdV--, EKdV-- and Boussinesq--related kink excitations,
i.e. $u_1$, $u_2$ and $u_3$,
are depicted in Figs. \ref{figure10} -- \ref{figure12}, for
three different values of $M$, $1.1$, $1.25$ and $2$.
We see that the EKdV and Bq models allow for both
compressive and rarefactive structures, while the KdV description
predicts a localized compression, which is quite sensitive to
velocity changes.
As expected (cf. the discussion above), both the EKdV-- and Bq--related
compressive excitations are similar in magnitude to the KdV--related anti-kink
for near--sonic velocity (i.e. near $M \approx 1$).
Nevertheless, we see that the KdV--related antikink becomes taller
and narrower as velocity increases, and substantially differentiates itself
from its EKdV-- and Bq--analogues.
One may wonder whether or not
the KdV picture (more familiar since widely studied)
is adequate for the modeling of a real
DP crystal, and also whether the rarefactive excitations predicted by other
theories can indeed be sustained in the crystal.
These questions may be answered by appropriate experiments and,
possibly, also be investigated by numerical simulations.
From a purely theoretical point of view, the Boussinesq--based description
appears to be more rigorous (recall that the KdV was derived in some
approximation) and valid in a more extended region than both the KdV
and Extended-KdV theories.

\section{Discreteness effects}

The above analytical solutions have been derived in the continuum
limit, i.e. for $L >> r_0$, where $L$ is the typical spatial
dimension (width) of the solitary excitation. One may therefore
define the discreteness parameter $g = r_0/L$, and require
\textit{a posteriori} that $g \ll 1$. From the
expressions derived for the Bq equation above, one easily sees that $g
\sim (v^2 - v_0^2)^{1/2}/v_1$, so this requirement is indeed
fulfilled for propagation velocities $v \approx v_0$. However, for
higher values of $v$, the (narrower) soliton will be subject to a
variety of effects e.g. shape distortion, wave radiation etc., due
to the intrinsic lattice discreteness. These effects have been
investigated in solid state physics \cite{Peyrard1986,
Flytzanis1989} and may be considered with respect to DP crystals
at a later stage. Let us briefly point out that narrow kink-shaped
lattice excitations have been numerically shown to propagate with
no considerable loss of energy, in a quite general monoatomic
lattice model \cite{Peyrard1986}.

Also worth mentioning is the
work of Rosenau \cite{Rosenau} who derived an improved version
of the Boussinesq equation (the I-Bq Eq.) in a quasi-continuum
limit. The I-Bq equation, which bears the general structure of
(\ref{GB}) upon replacing $h \, u_{xxxx}$ therein by $h \,
u_{xxtt}$ (yet with different coefficient definitions), is not
integrable and bears solitary wave solutions which do not collide
elastically; nevertheless, it was numerically shown to be more stable
than the Bq equation, and was argued to model discrete
lattice dynamics more efficiently, upon comparison of theoretical
predictions to exact numerical results \cite{Flytzanis1989}.
Further examination of such effects may be carried out in
dust-lattices, once our feedback from experiments has sufficiently
determined the relevance of the issue in real DP crystals, i.e.
typical excitation width, dynamics etc.

It should be underlined that the possibility for the
existence of breather
solitons, anticipated above, establishes a link between complex
plasma `solid state' modeling and the framework of
discreteness-related localized excitations (discrete breathers
\cite{breathers}, intrinsic localized modes \cite{Kiselev}), which
have recently received increasing interest among researchers in
the nonlinear dynamics community. These localized modes, which are
due to coupling anharmonicity and are stabilized by lattice
discreteness, have been shown to exist in frequency regions
forbidden to ordinary lattice waves and account for energy
localization in highly discrete real crystals, where continuum
theories fail. The relevance of this framework to dust crystals
appears to be an interesting open area for investigation.

\section{Envelope Excitations and shocks -- open issues}

As a final interesting issue involved in the nonlinear dynamics of
longitudinal lattice oscillations, let us
mention the nonlinear modulation of the amplitude of dust-lattice
waves, a well-known mechanism related to harmonic generation and,
possibly, the modulational instability of waves propagating in
lattices, eventually leading to modulated wave packet energy
localization via the formation of envelope solitons
\cite{Peyrard}. This framework, which was recently also
investigated with respect to low-frequency (dust-acoustic,
dust-ion acoustic) electrostatic waves in dusty plasmas
\cite{IKPKSPoP}, has been partly analyzed, on the basis of the
Melands\o \cite{Melandso} model in Ref. \cite{AMS2}. The authors
relied on a truncated Boussinesq equation, in the form of
(\ref{GB}) for $q = 0$, and succeeded in predicting the occurrence
of modulational instability in LDL waves in DP crystals and the
formation of envelope structures. Nevertheless, the nonlinearity
coefficient $q$ omitted therein seems to compete with $p$ in
(\ref{GB}) (notice the different signs) and is rather expected to
affect significantly the wave's stability profile.
It should be stressed that these localized
envelope excitations result from
a physical mechanism which is intrinsically
different from the one related to the small--amplitude excitations
described in this paper; see the discussion in Ref. \cite{Fedele}.
An
extended study of this modulation nonlinear mechanism is in row
and will be reported elsewhere, for clarity and
conciseness.

As a final comment, we may speculate on the role of damping,
herewith ignored, on the dynamics of dust-lattice waves. It is
known that weak damping may balance nonlinearity, leading to the
formation
of shock
wave fronts, as predicted in Refs. \cite{PKS2003, PKSAM} and
confirmed by numerical simulations \cite{Melandsoshocks}.
Furthermore, it was recently shown that the same mechanism may
result in the formation of large-amplitude wide-shaped solitary
waves, which may later break into a (gradually damped) train of
solitons or a wavepacket depending on physical parameters
\cite{Grimshaw}. We see that friction, yet weak, may play a
predominant role in the life and death of localized excitations;
this effect definitely deserves paying close attention with
respect to waves propagating in dust crystals. Again, one would
expect phenomenological theories followed by appropriately
designed experiments to elucidate the friction mechanisms inherent
in longitudinal dust-lattice wave propagation, in view of a more
complete description than the one provided by the conservative
model adopted here.

\section{Conclusions}

This work was devoted to an investigation of the relevance of
existing model nonlinear theories to the dynamics of longitudinal
oscillations in anharmonic chains, with emphasis on dust-lattice
excitations in (strongly-coupled) complex plasma crystals. Taking
into account an arbitrary interaction potential and long-range
interactions, we have rigorously shown that both compressive and
rarefactive kink-shaped (shock-like) excitations may form and
propagate in the lattice, depending basically on the mechanism of
interaction between grains located at each site. These excitations
are effectively modeled by either KdV- or Boussinesq-type
equations, whose analytic form was presented and whose qualitative
and quantitative differences were discussed. In any
case, the theory predicts coherent wave propagation above the
lattice's `sound' speed, in agreement with previous theoretical
works and experimental observations (in both atomic and
dust-lattices). It may be appropriate to mention that subsonic
soliton propagation in monoatomic chains was also numerically
considered and shown to be feasible in the past
\cite{Peyrard1986}. Let us point out that the model used here to
pass from a discrete description to the continuum
(long-wavelength) limit is quite generic, so possible modification
via refined nonlinear equations may readily be obtained from it,
for future consideration.

Furthermore, we have discussed the possibility of the formation of
breather modes and envelope excitations, as a consequence of
modulated wave packet instability, anticipating their link to
discrete nonlinear theories of localized modes, left for future
consideration; despite their analytical complexity, these models
may, in principle, be of relevance in dust crystals due to the
finite dimensions of the chain and its intrinsic spatial
discreteness. Finally, the possible role played by dissipation
mechanisms has been briefly discussed.

The present study relies on, and aims at extending, previous theories on
both anharmonic atomic chains and dusty plasma crystals. We hope
to have succeeded in reviewing the former (extending them to the
case of long-range electrostatic interactions) and generalizing
the latter (which are still in an early stage). Hopefully, our
predictions may be confirmed by appropriately set-up experiments,
with the ambition of throwing some light in the relatively new and
challenging field of strongly-coupled complex plasmas and
dust-lattice dynamics.

\begin{acknowledgments}
This work was supported by the European Commission (Brussels)
through the Human Potential Research and Training Network via the
project entitled: ``Complex Plasmas: The Science of Laboratory
Colloidal Plasmas and Mesospheric Charged Aerosols'' (Contract No.
HPRN-CT-2000-00140).
\end{acknowledgments}

\newpage

\begin{appendix}

\section{Computation of the coefficients for Debye (Yukawa) interactions}

Consider the Debye potential $\phi_D(r) = Q\, e^{-r/\lambda_D}/r$.
Defining the (positive real) lattice parameter $\kappa =
r_0/\lambda_D$, it is straightforward to evaluate the quantities
\[
\phi'_D(l r_0) = - \frac{Q}{\lambda_D^2} \, e^{-l \kappa}\,
\frac{1 + l \kappa}{(l \kappa)^2} \, , \]
\[ \phi''_D(l r_0) = +
\frac{2 Q}{\lambda_D^3}\, e^{-l \kappa}\,  \frac{1 + l \kappa+
\frac{(l \kappa)^2}{2}}{(l \kappa)^3} \, ,
\]
\[
\phi'''_D(l r_0) = - \frac{6 Q}{\lambda_D^4} \, e^{-l \kappa}\,
\frac{1 + l \kappa+ \frac{(l \kappa)^2}{2} + \frac{(l
\kappa)^3}{6}}{(l \kappa)^4} \, , \]
\[ \phi''''_D(l r_0) = +
\frac{24 Q}{\lambda_D^5} \, e^{-l \kappa}\, \frac{1 + l \kappa +
\frac{(l \kappa)^2}{2}+ \frac{(l \kappa)^3}{6} + \frac{(l
\kappa)^4}{24}}{(l \kappa)^5} \, ,
\]
where the prime denotes differentiation and $l = 1, 2, 3, ...$ is
a positive integer. Now, we shall combine these expressions
with Eqs. (\ref{defv0}), (\ref{defv1}), (\ref{defc11}) and
(\ref{defc111}), defining $v_0^2$, $v_1^2$, $p_0$ and $q_0$,
respectively.

Let us define the general (families of) sum(s)
\begin{eqnarray}
S_n(a) \, = \sum_{l = 1}^\infty a^l\, l^n \, \qquad \qquad  \hat
S_n^{(N)}(a) \, = \sum_{l = 1}^N a^l\, l^n \,
\nonumber \\
 (0 < a < 1) \, ,
\end{eqnarray}
(thinking of $a = e^{-\kappa}$, in particular); note that $\hat
S_n^{(N)}(a) \rightarrow S_n(a)$ for $ N \rightarrow \infty$;
also, $\hat S_n^{(1)}(a) = a$.
Making use of the well-known
geometrical series properties:
\begin{eqnarray}
S_0(a) \, = \sum_{l = 1}^\infty a^l\, = \frac{a}{1-a} \, \qquad
\hat S_0^{(N)} \, = \sum_{l = 1}^N a^l\, =  \frac{a (1 -
a^N )}{1-a} \nonumber \\
 (0 < a < 1) \, , \qquad
\end{eqnarray}
it is straightforward to derive $S_n$, $\hat S_n^{(N)}$ for $l \ge
1$, by differentiating with respect to $a$. One obtains
\[
S_1(a) \, = \,\sum_{l = 1}^\infty a^l\,l = a \,\sum_{l = 1}^\infty
l \,a^{l-1} = \,a \, \sum_{l = 1}^\infty \frac{\partial
(a^l)}{\partial a}\,\]
\[= \, a \, \frac{\partial }{\partial a} \sum_{l =
1}^\infty a^l \, = \, a \, \frac{\partial S_0}{\partial a} \, =
\frac{a}{(1-a)^2} \, .
\]
In a similar manner, iterating from
\[{\partial^2 (a^l)}/{\partial
a^2} = \,l \,(l-1) \, a^{l-2}\, = \, a^{-2}\,(l^2 a^{l}- l
a^{l}) \, ,\]
one finds
\[
S_2(a) \, = \,\biggl( a^2 \, \frac{\partial^2 }{\partial a^2} + a
\, \frac{\partial }{\partial a} \biggr) \, S_0 \, = \, ... \, =
\frac{a (1+a)}{(1-a)^3} \, ;
\]
then
\[
S_3(a) \, = \,\biggl( a^3 \, \frac{\partial^3 }{\partial a^3} + 3
a^2 \, \frac{\partial^2 }{\partial a^2} + a \, \frac{\partial
}{\partial a} \biggr) \, S_0 \, \]
\[ = \, ... \, = \frac{a (a^2 + 4 a +
1 )}{(1-a)^4} \, ,
\]
and so forth. Also note the identity
\begin{equation}
S_{-1}(a) \, = \sum_{l = 1}^\infty \frac{a^l}{l} \, =  - ln(1-a)
\qquad \qquad \qquad
 (0 < a < 1) \, .
\end{equation}
The corresponding set of formulae may be obtained for $\hat
S_n^{(N)}$ in a similar manner.

Now, substituting $a = e^{-\kappa}$ and using the derivatives
of $\phi_D$ above, one may immediately evaluate the expressions
(\ref{defv0}), (\ref{defv1}), (\ref{defc11}) and (\ref{defc111}).
Setting $r_0 = \kappa \lambda_D$ everywhere, it is straightforward
to show that
\begin{eqnarray} c_2 \equiv v_0^2 \, \equiv
\, \omega_{0, L}^2 \,r_0^2 =  \frac{Q}{M} \,\kappa^{2}
\,\lambda_D^2 \sum_{l = 1}^\infty \, l^2 \, \phi''(l r_0) \, =\,
... \qquad \nonumber \\ = \, \frac{2 Q^2}{M \lambda_D} \, \biggl[
\kappa^{-1} \, S_{-1}(e^{-\kappa})\, + \, \kappa^0 \,
S_0(e^{-\kappa})\, + \, \frac{1}{2} \kappa^1 \, S_1(e^{-\kappa})
\biggr] \, .\nonumber \\
 \label{defv0-Debye}
\end{eqnarray}
In the same manner
\begin{eqnarray} \frac{c_4}{r_0^2} \equiv v_1^2 \,=
\frac{Q}{12 M} \,\kappa^{2} \,\lambda_D^2 \,
 \sum_{l = 1}^\infty \,
l^4 \, \phi''(l r_0) =\, ... \,\nonumber \\
=\, \frac{Q^2}{6 M \lambda_D} \, \biggl[ \kappa^{-1} \,
S_1(e^{-\kappa})\, + \, \kappa^0 \, S_2(e^{-\kappa})\, \nonumber \\
+ \,
\frac{1}{2} \kappa^1 \, S_3(e^{-\kappa}) \biggr] \, .
 \label{defv1-Debye}
\end{eqnarray}
Also
\begin{eqnarray} p_0 \equiv \, - c_{11} \,
 =  -  \frac{Q}{M} \,\kappa^{3} \,\lambda_D^3 \sum_{l =
1}^\infty \, l^3 \, \phi'''(l r_0) \, \nonumber \\ = \, ... = \,
\frac{6 Q^2}{M \lambda_D} \, \times \, \nonumber \\
\biggl[ \kappa^{-1} \,
S_{-1}(e^{-\kappa})\, + \, \kappa^0 \, S_0(e
\nonumber \\
^{-\kappa})\, + \,
\frac{1}{2} \kappa^1 \, S_1(e^{-\kappa}) \, + \, \frac{1}{6}
\kappa^2 \, S_2(e^{-\kappa}) \biggr] \, .
 \label{defc11-Debye}
\end{eqnarray}
Finally, from (\ref{defc111}) we have
\begin{eqnarray}
q_0 \equiv \, c_{111} \,
= \frac{Q}{2 M} \,\kappa^{4} \,\lambda_D^4 \sum_{l = 1}^\infty \,
l^4 \, \phi''''(l r_0) \, =\, ... \qquad \qquad  \nonumber
\\ = \, \frac{12 Q^2}{M \lambda_D} \, \biggl[  \kappa^{-1} \,
S_{-1}(e^{-\kappa})\, + \, \kappa^0 \, S_0(e^{-\kappa})\,
\qquad \qquad \qquad
\nonumber
\\
+ \,
\frac{1}{2} \kappa^1 \, S_1(e^{-\kappa}) \, + \,
\frac{1}{6}
\kappa^2 \, S_2(e^{-\kappa}) \, + \, \frac{1}{24} \kappa^3 \,
S_3(e^{-\kappa}) \biggr] \, . \qquad
 \label{defc111-Debye}
\end{eqnarray}
The corresponding expressions for a value of $N$ are given by
substituting $S_n(\cdot)$ with $\hat S_n^{(N)}(\cdot)$ everywhere.
One immediately sees that $p_0/v_0^2 > 2$, $q_0/v_0^2 > 6$; also,
$v_1^2/v_0^2 = 12$ for $N=1$ (only), i.e. for the NNI case.

Finally, combining the above exact expressions for
$S_{-1}(a)$, ..., $S_{3}(a)$, we obtain exactly expressions
(\ref{Ninfinite-omega0}) to (\ref{Ninfinite-q0}) in the text.

\end{appendix}

\newpage


\centerline{\textbf{Figure Captions}}

\bigskip

Figure 1.

\smallskip

(a) The linear oscillation frequency squared $\omega^2$
(normalized over $Q^2/(M \lambda_D^3)$) is depicted against the
lattice constant $\kappa$, for $N=1$ (first-neighbor
interactions: ---), $N=2$ (second-neighbor interactions: - - -), $N=\infty$
(infinite-neighbors: -- -- --), from bottom to top.
(b) Detail near $\kappa
\approx 1$.

\bigskip

Figure 2.

\smallskip

(a) The characteristic 2nd order dispersion velocity squared
$v_0^2$ (normalized over $Q^2/(M \lambda_D)$) is depicted against
the lattice constant $\kappa$, for $N=1$ (first-neighbor
interactions: ---), $N=2$ (second-neighbor interactions: - - -), $N=\infty$
(infinite-neighbors: -- -- --), from bottom to top. (b) Detail near $\kappa
\approx 1$.

\bigskip

Figure 3.

\smallskip

(a) The characteristic 4th order dispersion velocity squared
$v_1^2$ (normalized over $Q^2/(M \lambda_D)$) is depicted against
the lattice constant $\kappa$, for $N=1$ (first-neighbor
interactions: ---), $N=2$ (second-neighbor interactions: - - -), $N=\infty$
(infinite-neighbors: -- -- --), from bottom to top. (b) Detail near $\kappa
\approx 1$.

\bigskip

Figure 4.

\smallskip

(a) The nonlinearity coefficient $p_0$ (normalized over $Q^2/(M
\lambda_D)$) is depicted against the lattice constant $\kappa$
for $N=1$ (first-neighbor interactions: ---), $N=2$
(second-neighbor interactions: - - -), $N=\infty$
(infinite-neighbors: -- -- --), from bottom to top. (b) Detail near $\kappa
\approx 1$.

\bigskip

Figure 5.

\smallskip

(a) The nonlinearity coefficient $q_0$ (normalized over $Q^2/(M
\lambda_D)$) is depicted against the lattice constant $\kappa$
for $N=1$ (first-neighbor
interactions: ---), $N=2$ (second-neighbor interactions: - - -), $N=\infty$
(infinite-neighbors: -- -- --), from bottom to top. (b) Detail near $\kappa
\approx 1$.

\bigskip

Figure 6.

\smallskip

Dispersion relation for the Debye interactions, neglecting damping;
cf. (\ref{dispersion-Debye}) for $\nu = 0$:
the square frequency $\omega^2$, normalized over
${Q^2}/({M \lambda_D^3})$, is depicted
versus the normalized wavenumber $k r_0/\pi$ for $N=1$ (first-neighbor
interactions: ---), $N=2$ (second-neighbor interactions: - - -), $N = 7$
(up to 7th nearest-neighbors: -- -- --), i.e. from bottom to top.

\bigskip

Figure 7.

\smallskip

Localized antikink/kink (negative/positive pulse) functions,
related to the KdV Eq. (\ref{KdV}), for the displacement $u(x, t)$
(relative displacement $w(x, t) \sim \partial u(x, t)/\partial
x$), for positive/negative $p_0$ coefficient i.e. $s = +1/-1$, are
depicted in figures (a)/(b); recall that (a) holds for Debye
interactions; arbitrary parameter values: $v = 1$ (solid curve),
$v = 2$ (long dashed curve), $v = 3$ (short dashed curve).

\bigskip

Figure 8.

\smallskip

(a) The two localized pulse solutions of the EKdV Eq. (\ref{EKdV})
for the relative displacement $w(x, t) \sim \partial u(x,
t)/\partial x$ are depicted for some set of (positive) values of
the $p_0$ and $q_0$ coefficients (i.e. $s = +1$): the first
(dashed curve)/second (short--dashed) solution, as given by
(\ref{EKdVsol1})/(\ref{EKdVsol2}), represents the smaller
negative/larger positive pulses. The larger negative pulse (solid
curve) denotes the solution of the KdV Eq. (\ref{KdV}) for the
same parameter set. (b) The corresponding solutions for the
particle displacement $u(x, t)$.

\bigskip

Figure 9.

\smallskip

(a) The (normalized) maximum value of the kink--shaped localized
displacement $u_1(x, t)/r_0$, as obtained from the KdV Equation,
is depicted versus the lattice parameter $\kappa$ and the
normalized velocity (Mach number) $M = v/v_0$. (b) The
(normalized) width $L_1/r_0$ of $u_1(x, t)$ is depicted against $M
= v/v_0$.

\bigskip

Figure 10.

\smallskip

The antikink excitation predicted by the KdV theory (solid curve)
is compared to the (two) solutions obtained from
(a) the EkDV Equation;
(b) the Bq Equation--related model (dashed curves).
Values: lattice parameter $\kappa = 1.1$,
normalized velocity (Mach number) $M = v/v_0 = 1.25$.

\bigskip

Figure 11.

\smallskip

\begin{figure}[htb]
 \centering
 \resizebox{3in}{!}{
 \includegraphics[]{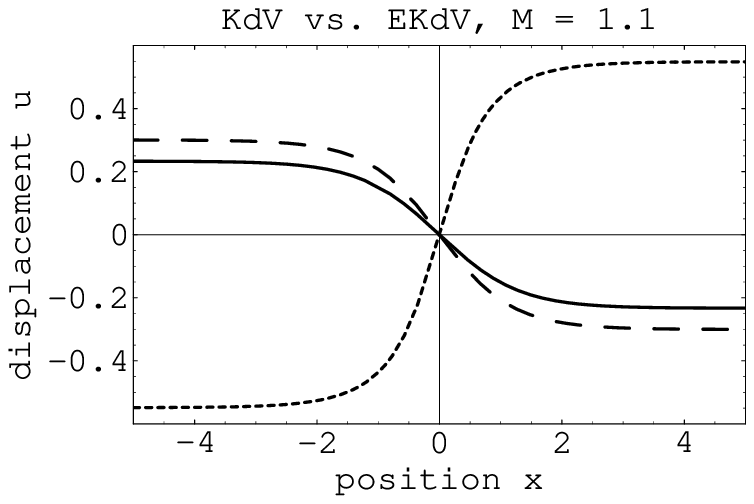}} \\
\vskip 2 cm \resizebox{3in}{!}{
 \includegraphics[]{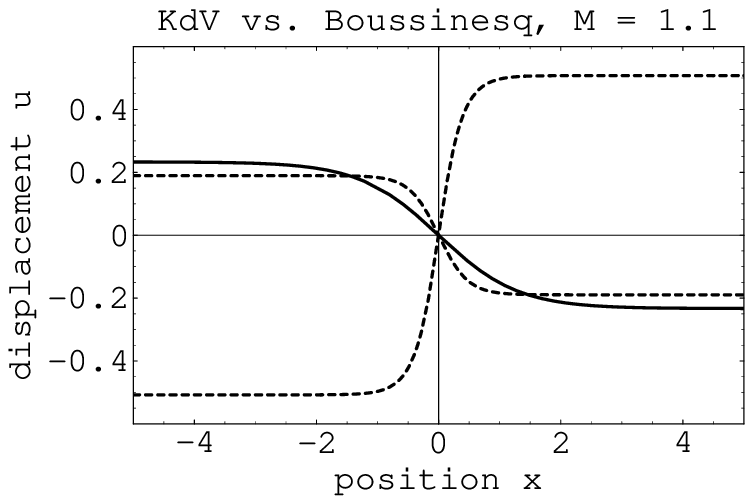}}
\caption{} \label{figure10}
\end{figure}

Similar to Fig. \ref{figure10}, for $M = v/v_0 = 1.25$.

\bigskip

Figure 12.

\smallskip

\begin{figure}[htb]
 \centering
 \resizebox{3in}{!}{
 \includegraphics[]{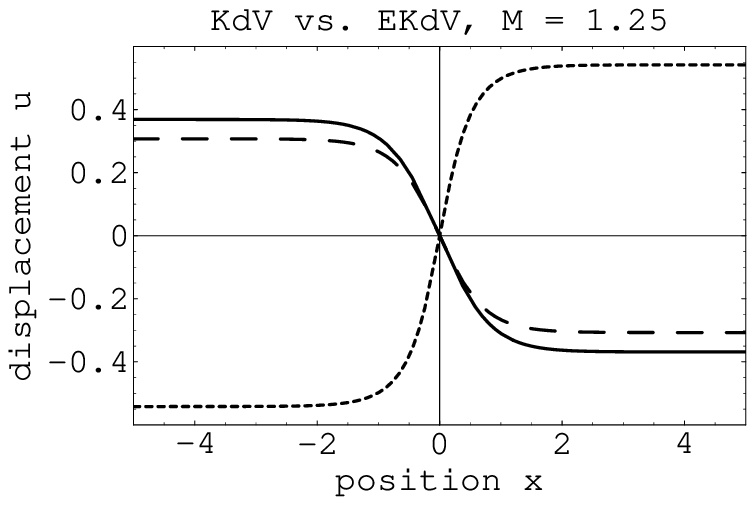}} \\
\vskip 2 cm \resizebox{3in}{!}{
 \includegraphics[]{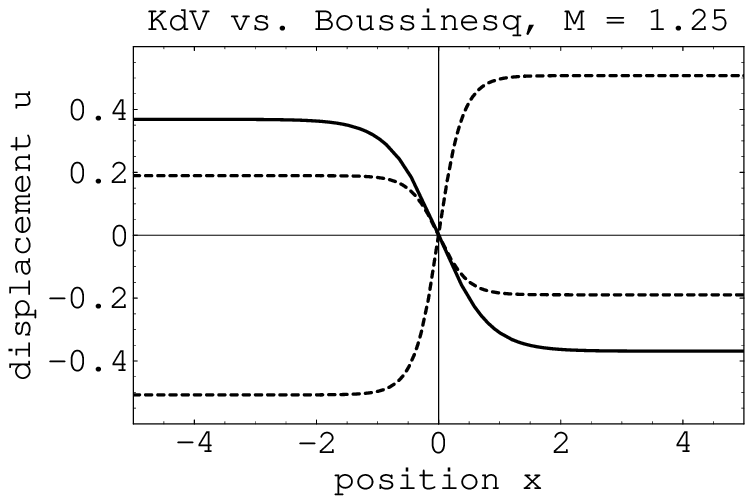}}
\caption{} \label{figure11}
\end{figure}

\begin{figure}[htb]
 \centering
 \resizebox{3in}{!}{
 \includegraphics[]{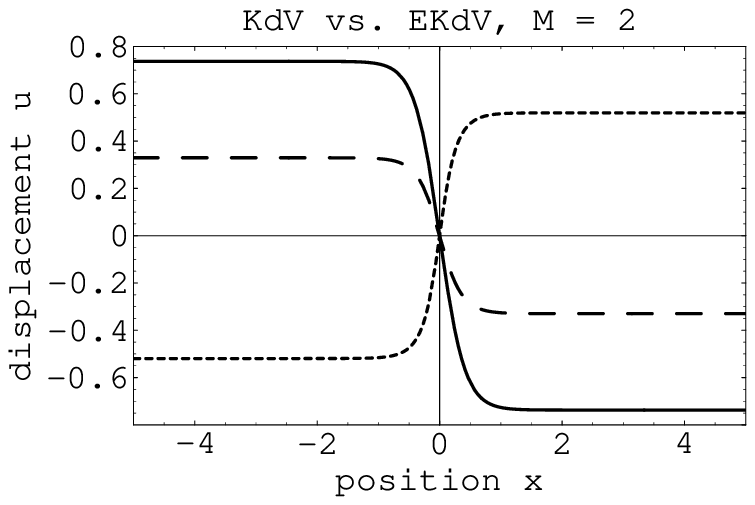}} \\
\vskip 2 cm \resizebox{3in}{!}{
 \includegraphics[]{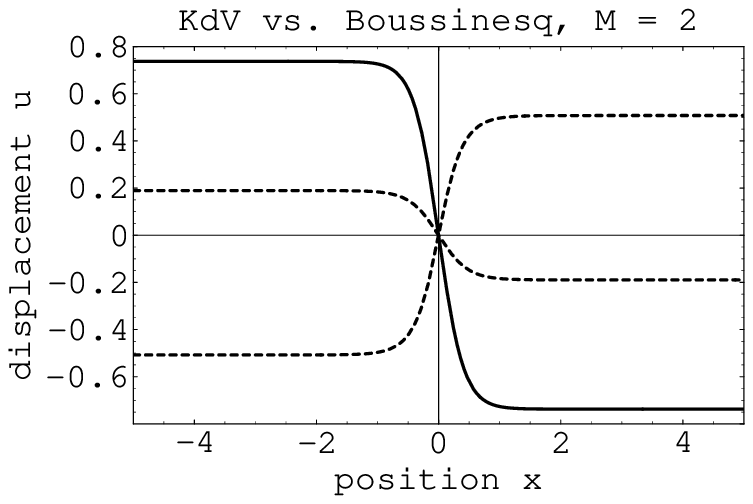}}
\caption{} \label{figure12}
\end{figure}

Similar to Figs. \ref{figure11} and \ref{figure12}, for $M = v/v_0 = 2$.

\bigskip

\newpage

\newpage

\vskip 2 cm

\newpage

\vskip 2 cm

\newpage

\vskip 2 cm

\newpage

\vskip 2 cm

\newpage

\vskip 2 cm

\newpage

\vskip 2 cm

\newpage

\vskip 2 cm

\newpage

\vskip 2 cm

\newpage

\vskip 2 cm

\newpage

\vskip 2 cm

\newpage

\vskip 2 cm

\begin{thebibliography}{10}
\bibitem{commentWadati1} In order to see this, transform the variables in
Ref. \cite{Wadati} as: $u \rightarrow w$, $x \rightarrow \zeta/\sqrt{b}$,
$t \rightarrow \tau/\sqrt{b}$,
$\alpha \rightarrow a/6$, $\beta \rightarrow \hat a/6$,
and then insert  a factor $(- s)$, since
formulae therein for $u$ correspond to $s = -1$ and,
upon $u \rightarrow - u$, to $s = -1$ here.
\end{thebibliography}
\end{document}